\documentclass[conference]{IEEEtran}
\usepackage{amsmath,amssymb,amsfonts}
\usepackage{graphicx}
\usepackage{textcomp}
\usepackage{xcolor}
\usepackage{acronym}
\usepackage{verbatim}
\usepackage{multirow}
\usepackage{makecell}
\usepackage{todonotes}
\usepackage{balance}
\usepackage{url}
\usepackage{pifont}
\usepackage{array}
\usepackage{adjustbox}
\usepackage{multirow}
\usepackage{tabularx}
\usepackage{amsmath}
\usepackage{color,soul}
\usepackage[ruled]{algorithm2e}
\usepackage{pdfpages}
\usepackage{caption}

\usepackage{pifont}
\newcommand{\cmark}{\ding{51}}%
\newcommand{\xmark}{\ding{55}}

\newcolumntype{M}[1]{>{\centering\arraybackslash}m{#1}}

\acrodef{Adam}{Adaptive Moment Estimation}
\acrodef{AoA}{Angle of Arrival}
\acrodef{AR}{AutoRegressive}
\acrodef{CNN}{Convolutional Neural Network}
\acrodef{DoA}{Direction of Arrival}
\acrodef{GMM}{Gaussian Mixtures Model}
\acrodef{HMM}{Hidden Markov Model}
\acrodef{LS-LDA}{Least Squares Linear Discriminant Analysis}
\acrodef{MFCC}{Mel-Frequency Cepstrum Coefficient}
\acrodef{PLP}{Perceptual Linear Prediction Coefficients}
\acrodef{PSD}{Power Spectral Density}
\acrodef{ReLU}{Rectified Linear Units}
\acrodef{RMSProp}{Root Mean Square Propagation}
\acrodef{SGD}{Stochastic Gradient Descent}
\acrodef{SNR}{Signal-to-Noise Ratio}
\acrodef{SVM}{Support Vector Machine}
\acrodef{STFT}{Short-Time Fourier Transform}
\acrodef{ToA}{Time of Arrival}
\acrodef{TDoA}{Time Difference of Arrival}
\acrodef{WASN}{Wireless Acoustic Sensor Network}
\acrodef{WSN}{Wireless Sensor Network}
\acrodef{ZCR}{Zero Crossing Rate}

\begin{document}

\title{Sound of Guns:\\Digital Forensics of Gun Audio Samples meets Artificial Intelligence}

\author{
    \IEEEauthorblockN{Simone Raponi, Gabriele Oligeri, Isra M. Ali} \\
    \IEEEauthorblockA{Division of Information and Computing Technology \protect\\ College of Science and Engineering, Hamad Bin Khalifa University - Doha, Qatar
    \\\{sraponi, goligeri, isali\}@hbku.edu.qa}
}

\maketitle


\begin{abstract}
    Classifying a weapon based on its muzzle blast is a challenging task that has significant applications in various security and military fields. Most of the existing works rely on ad-hoc deployment of spatially diverse microphone sensors to capture multiple replicas of the same gunshot, which enables accurate detection and identification of the acoustic source. However, carefully controlled setups are difficult to obtain in scenarios such as crime scene forensics, making the aforementioned techniques inapplicable and impractical. 
    
    We introduce a novel technique that requires zero knowledge about the recording setup and is completely agnostic to the relative positions of both the microphone and shooter. Our solution can identify the category, caliber, and model of the gun, reaching over 90\% accuracy on a dataset composed of 3655 samples that are extracted from YouTube videos. 
    Our results demonstrate the effectiveness and efficiency of applying Convolutional Neural Network (CNN) in gunshot classification eliminating the need for an ad-hoc setup while significantly improving the classification performance.
\end{abstract}

\IEEEpeerreviewmaketitle

\section{Introduction}
\label{sec:introduction}

Gunshot analysis have received significant attention from both the military and scientific communities. Acoustic analysis of gunshots can provide useful information, such as the position of the shooter, the projectile trajectory, the caliber of the gun, and the gun model. Although acoustical evidence may significantly contribute to audio forensic reconstruction and analysis, the forensic analysis of gunshots is characterized by many challenges due to the broadcast and noisy nature of the acoustic channel. 

Consider a scenario where a microphone is deployed in a close neighborhood to the shooter. The recorded audio sample can be significantly affected by the environmental surroundings, such as trees, foliage, and buildings, which attenuate and reflect the main component of the shock wave. The resulting audio sample may feature different echoes of the gunshot that are characterized by different attenuation factors as a function of their paths. This naive approach is impractical, which motivated the development of more complex ad-hoc acoustic data acquisition strategies over the last decade. 
To mitigate echoes and overcome the intrinsic lack of information, that the aforementioned scenario suffers from, additional microphones are deployed. The comparison of multiple replicas of the same gunshot enables \emph{shooter localization} and \emph{weapon identification}. The physical characteristics of acoustic propagation can be exploited to infer the position of the shooter and the category of the gun. Multiple spatially diverse acoustic sources enable the estimation of \ac{AoA}, \ac{ToA}, and \ac{TDoA}. 
 The obtained recordings can be modeled by geometrical acoustics that enable the localization of the shooter.
 Furthermore, multiple replicas of the same acoustic source allow to filter out echoes and background noise affecting a subset of the deployed microphones, thus enabling a deep characterization for both the time and frequency domains.

Acoustic acquisition via \ac{WSN} 
requires a specialized infrastructure overlay to enable sensor communication, data processing, and computation distribution. Solutions that rely on the spatial diversity provided by the WSN 
introduce several types of burdens. Firstly, each soldier has to carry a wearable device equipped with a microphone and other sensors, such as a compass, to collect meaningful information about \ac{AoA}, \ac{ToA}, and \ac{TDoA}. Secondly, in a military scenario, the \ac{WSN} should feature a jamming-resistant communication protocol and non-interfering radio channels. Both assumptions are difficult to achieve given the resource constraints of \ac{WSN}s in terms of CPU, battery, and memory. In most cases, \ac{WSN}s cannot afford the computational burden of multimedia processing. Therefore, the captured data should be first off-loaded to a remote server, then downloaded and distributed again. This represents a challenge from the connectivity perspective since, in many cases, military \ac{WSN}s are unattended or provided with a discontinued link to the control center.


In this work, we do not rely on ad-hoc acquisition setups, but we exploit publicly available audio recordings of gunshots, considering their temporal and spectral representations. Spectral analysis of sound has been adopted in many contexts to detect and identify recurrent patterns. In particular, the combination of time-frequency decomposition of audio samples with \ac{CNN} provides promising performance in detecting recurrent patterns~\cite{SALEEM2020300982}. The \ac{CNN} is trained over several ``images'' constituted by a three-dimensional representation of time, frequency, and amplitude. The result is a robust solution that can ``recognize'' the same sound by cross-matching similar images.

{\bf Contribution.} We propose an inexpensive solution that is able to detect and identify gunshots without resorting to any ad-hoc infrastructure. Contrary to other studies, our solution requires only an audio sample of a gunshot that can be easily obtained by any commercially available microphone. Our approach is agnostic to the microphone position with respect to the shooter, and it does not require multiple spatially different replicas of the gunshot; we consider recordings from mono-channel setups with different sample rates. We proved the effectiveness of our solution by considering 3655 samples of gunshots constituted by 30 pistols, 18 rifles, and 11 shotguns for a total of 7 different calibers. The proposed approach guarantees an accuracy higher than 90\% for all of the considered cases, namely, the category, model and caliber of the gun. 

{\bf Paper organization.} The remainder of this paper is organized as follows. Section~\ref{sec:related_work} summarizes recent contributions in the field of weapon classification. Section~\ref{sec:background} introduces the background concepts related to frequency domain analysis, \ac{CNN}s, and acoustic characteristics of gunshots. Section~\ref{sec:dataset_description} describes our dataset and Section~\ref{sec:dataset_generation} discusses the the dataset generation process. The neural network architecture is presented in Section~\ref{sec:overall_architecture}. Section~\ref{sec:performance} shows the performance of our solution. Finally, Section~\ref{sec:conclusion} draws some concluding remarks.

\section{Related Work}
\label{sec:related_work}
Firearm classification based on the acoustic evidence generated by its discharge has long been investigated, but not extensively studied in the literature. Proposed solutions vary in many aspects, including the source of acoustic data, the type of analysis applied, the type of features extracted, and the application area. Table~\ref{table:relatedWork} summarizes prior studies, that provide gunshot classification and firearm identification, according to these aspects.

The source of the data is characterized by the type, the quality, and the environmental conditions of the deployed audio recording setup, which defines the amount of information that can be leveraged for classification. Most of the gunshot recordings used in the literature are either obtained under carefully controlled conditions, where a distributed set of microphone sensors are deployed~\cite{sallai2011weapon, libal2014wavelet, sanchez2017maximum}, or extracted from a conventional recording device in less controlled environments~\cite{khan2010weapon, khan2009weapon, djeddou2013classification, morton2011classification, kiktova2015gun}. 

In the former case, where a \ac{WASN} is deployed, spatial information can be obtained by performing array processing and triangulation techniques. \ac{DoA} and \ac{ToA} estimation methods are applied to the obtained audio signals to determine the projectile speed and trajectory, as well as to infer the position of the shooter. Such information may also provide discriminant features, such as the bullet speed~\cite{sallai2011weapon}, that can be used to identify the firearm category. Furthermore, the distributed nature of the recording setup provides spatial diversity, where multiple acoustic observations from different locations of the same gunshot are obtained, which can be leveraged to increase the classification accuracy. S{\'a}nchez-Hevia et al.~\cite{sanchez2017maximum} exploited this feature and proposed a multi-observation weapon classification system that leverages various classifier ensembles to enhance classic decision fusion techniques. Each node in the sensor network produces a classification decision using \ac{LS-LDA}. The decisions are later fused using a Maximum Likelihood-based fusion rule that weights the decision of each node based on its location. 

The main constraint induced by this type of analysis is the requirement of spatial information, which can only be obtained by deploying a distributed sensor network. Therefore, limiting the applicability of gunshot detection and firearm classification to a carefully controlled recording setup only. Consequently, various pattern recognition approaches were proposed that identify the firearm category in the absence of spatial information. The most used classifiers for firearm identification are \ac{GMM}~\cite{khan2010weapon, khan2009weapon, djeddou2013classification} and \ac{HMM}~\cite{morton2011classification, kiktova2015gun}. 

Most of these approaches can be described as frame-based feature classification approaches~\cite{khan2010weapon, khan2009weapon, djeddou2013classification, kiktova2015gun}, where the time-domain acoustic signal is subdivided into a sequence of short-time windowed frames. From each frame, a set of predetermined features is extracted and used for gunshot classification. The most common extracted features are statistical measures of the spectrum and intensity of the signal, in addition to perceptual features such as \ac{MFCC} or \ac{PLP}. Temporal features, such as energy and \ac{ZCR}, are also used, but only in conjunction with spectral or perceptual features.

\begin{table*}[]
\centering
\caption{Prior Gunshot Classification Approaches}
    \begin{adjustbox}{width=\textwidth}
        \begin{tabular}{|M{1cm}|M{3.7cm}|M{1.4cm}|M{2.2cm}|M{2.3cm}|M{1.3cm}|M{1.4cm}|}
        \hline
        \textbf{Name} & \textbf{Technique/Classifier} & \textbf{Features} & \textbf{Dataset} & \textbf{Result} & \textbf{Varying conditions} & \textbf{No Ad-Hoc Setup} \\ \hline
        \cite{khan2009weapon} & Hierarchical GMM classification & Cepstral & 50-100 gunshots of 10 gun types & 90\% (category) 85\% (caliber) & \xmark & \cmark \\ \hline
        \cite{khan2010weapon} & Exemplar embedding using hierarchical GMM classification & Cepstral & 100 gunshots of 20 gun types & 95-100\% (category) 60-72\% (model) & \cmark  & \xmark \\ \hline
        \cite{sallai2011weapon} & DoA \& ToA estimation & Projectile speed & 194 gunshots of 4 gun types & 86\% (caliber) & \textbf{-} & \xmark \\ \hline
        \cite{morton2011classification} & HMM using Non-parametric Bayesian techniques & None & $\sim$46 gunshots of 5 gun types & 95.65\% (model) & \xmark & \cmark  \\ \hline
        \cite{djeddou2013classification} & Hierarchical GMM classification & Cepstral \& Temporal & 230 gunshots of 5 gun types & 96.29\% (category) & \xmark & \cmark \\ \hline
        \cite{kiktova2015gun} & HMM classification \& Viterbi based decoding & Spectral \& Temporal & 372 gunshots of 4 handgun types & 80\% (model) & \textbf{-} & \cmark \\ \hline
        \cite{sanchez2017maximum} & LS-LDA \& Maximum Likelihood decision fusion & Cepstral, Spectral \& Temporal & 840 gunshots of 14 gun types & 94.1\% (model)  & \cmark & \xmark \\ \hline
        {\bf Our solution} & {\bf Convolutional Neural Networks} & {\bf Spectral \& Temporal} & {\bf 3655 gunshots of 59 gun types (pistols, rifles, and shotguns)} & {\bf 90\% for Category, Caliber, and Model} & \cmark & \cmark  \\ \hline 
        \end{tabular}
        \label{table:relatedWork}
    \end{adjustbox}
\end{table*}


Morton et al.~\cite{morton2011classification} proposed an alternative classification approach that does not rely on frame-based features aiming to eliminate the dependency on performance-driven parameters, which are often optimized over a finite training set. They proposed modeling each firearm category as an \ac{HMM} with \ac{AR} source densities using non-parametric Bayesian priors to allow automated model order selection. The \ac{AR} defines a set of energy and spectral characteristics of the captured gunshot, while the \ac{HMM} identifies the transitions of these states. 

The aforementioned techniques may perform adequately in matched experimental conditions, however, their effectiveness could reduce significantly when capture conditions vary in challenging unstructured environments, where noise and distortion are present. 
Although Khan et al.~\cite{khan2010weapon} addressed this problem by using an exemplary embedding approach to bridge between varying recording conditions, the achieved classification accuracy is relatively low (i.e., 60-72\%). The authors used a dataset of 100 gunshot samples obtained from 20 different firearm models, where each model is represented by 5 to 15 gunshot samples. The different conditions included in their experiments were simulated, namely, ``Room Reverb'', ``Concert Reverb'', and ``Doppler Effect'', which may not match real-life environmental conditions and do not include directional variations. Furthermore, their approach assumes prior knowledge of the recording conditions which is not always possible, especially in audio forensic reconstruction analysis.

Our solution, being the only one considering varying environment conditions and not requiring an ad-hoc setup, outperforms the state of the art studies in terms of dataset richness, including the number of gunshots samples and range of weapon models, reaching 90\% accuracy.
\section{Background}
\label{sec:background}

\subsection{Spectrogram}

\begin{figure}[htbp]
    \includegraphics[width=\columnwidth]{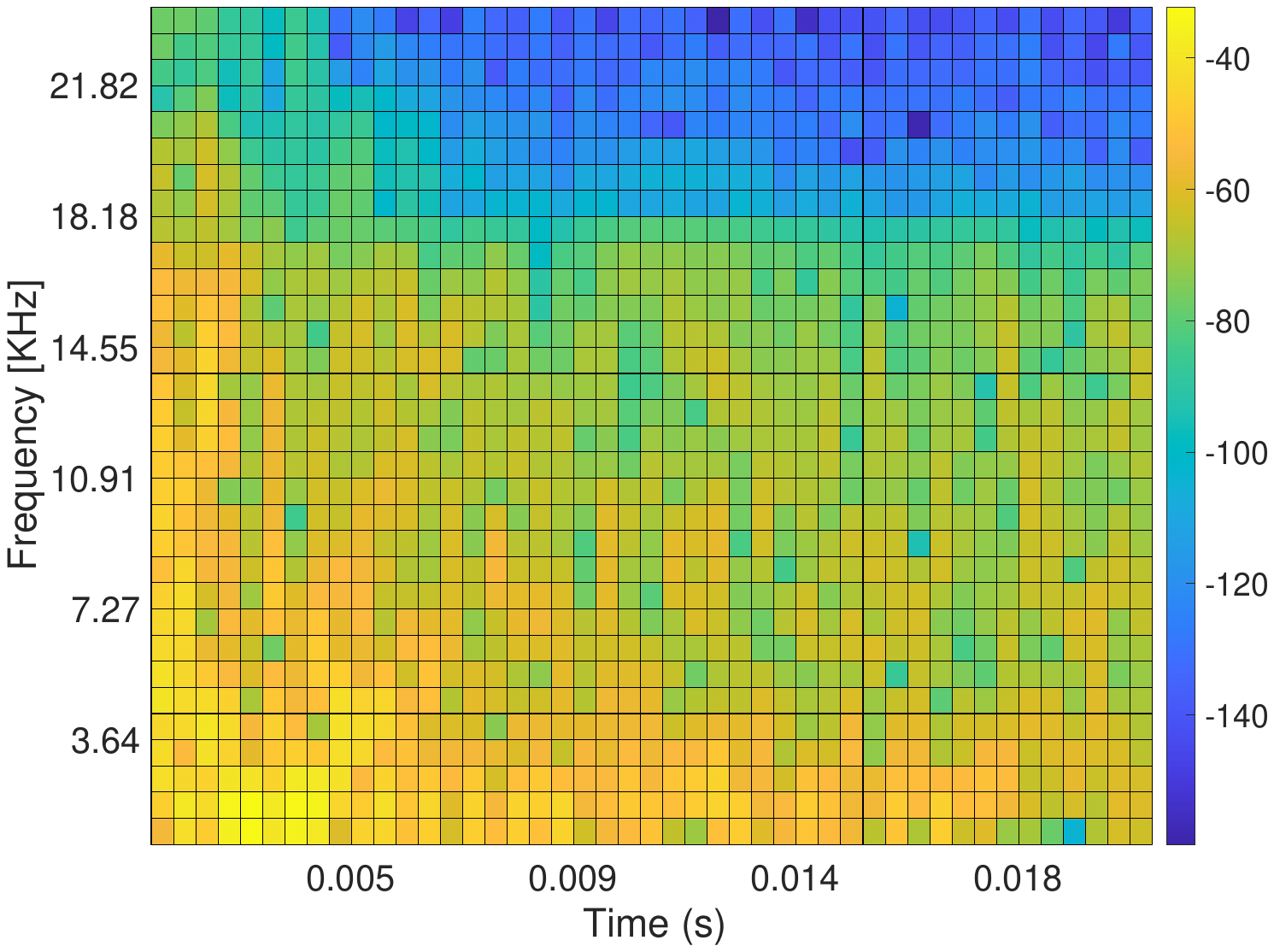}
    \centering
    \caption{Example of a gunshot spectrogram. The x-axis represents the time expressed in seconds, while the y-axis represents the frequencies expressed in kHz. The color represents the \ac{PSD} at the given time-frequency.}
    \label{fig:spectrogram}
\end{figure}

A spectrogram is one of the most widely adopted visual representations of the frequencies spectrum of a signal over time. Being defined as an intensity plot of the \ac{STFT} magnitude, a spectrogram is usually portrayed as a bi-dimensional graph, where one axis (usually the x-axis) represents time and the other axis (usually the y-axis) represents frequencies. An example of spectrogram is depicted in Fig.~\ref{fig:spectrogram}.
Each intersection between time and frequency is assigned a color that refers to the \ac{PSD} of that specific frequency at that particular time, 
which is considered a third dimension of the graph. 
To compute the spectrogram of a signal \textit{y}, the signal is divided into shorter fixed-length segments $y_1, \dots, y_n$, and the Fourier transform is applied separately to each segment. The spectrogram describes the changes of the signal frequencies spectrum as a function of time. This implies that, if the time is discrete, the data to be transformed may be partitioned into overlapping frames. The \ac{STFT} is applied to each of the frames and the result, consisting of both phase and magnitude for each intersection between time and frequency, is stored in a matrix, as showed in Equation~\ref{equation:spectrogram}.

\begin{equation}\label{equation:STFT}
    STFT\{y_n\}(m,\omega) = \sum_{n=-\infty}^{\infty} y_n\omega[n-m]e^{-j\omega n}
\end{equation}
\begin{equation}\label{equation:spectrogram}
    spectrogram\{y_n\}(m,\omega) = |STFT\{y_n\}(m,\omega)|^2
\end{equation}

The result consists of a bi-dimensional matrix that maps the audio frequencies to the time-localized points~\cite{ravi}.

The visual representation of audio traces through spectrograms have been extensively leveraged in the literature in the context of audio classification~\cite{hershey}, sound event classification~\cite{dennis}, emotion recognition~\cite{satt}, human activity recognition~\cite{ravi}, cross-modality feature learning~\cite{ngiam}, and gunshot classification~\cite{navratil}.

\subsection{Convolutional Neural Network}
\label{sec:cnn}

A \ac{CNN} belongs to the class of deep neural networks that have one or more convolutional layers (i.e., layers that perform convolution operations)~\cite{krizhevsky}. A convolution is a linear operation that consists of a slide of a parametric-sized filter over the input representation (usually a visual image). The application of the same filter to different overlapping filter-sized portions of the input generates a feature map. There are several types of filters, also known as operators. Each filter tries to identify a specific feature within the input representation. For example, the Sobel, the Prewitt, and the Canny operators highlight edges, the Harris and the Shi and Tomasi operators highlight corners, etc. One of the most powerful features of \ac{CNN}s, that is also the reason behind their wide adoption, is the ability to automatically apply an extensive number of filters to the input representation in parallel, thus highlighting specific features in every part of the input image simultaneously.

\ac{CNN}s can be seen as regularized versions, that discourage learning complex models, of multilayer perceptrons. While in multilayer perceptrons several fully connected layers are used---a layer is fully connected if all the neurons it is composed of are connected to all the neurons of the next layer, \ac{CNN}s exploit a hierarchical structure that allows building complex patterns by using small and simple patterns. 

\begin{figure}[htbp]
    \includegraphics[width=\columnwidth]{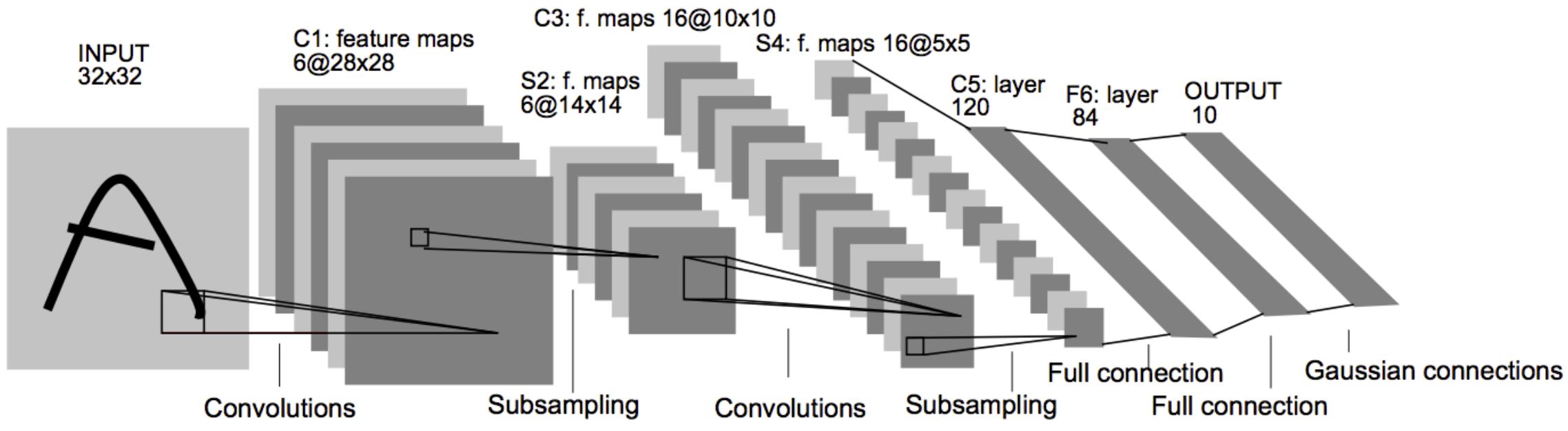}
    \centering
    \caption{Example of a \ac{CNN}. LeNet-5~\cite{lecun} is able to identify handwritten digits for zip code recognition in the postal service}
    \label{fig:lenet5}
\end{figure}

Fig.~\ref{fig:lenet5} depicts a typical architecture of a \ac{CNN}. The dimension of the input image (in this case representing a handwritten digit), keeps decreasing while going deeper in the neural network, while the number of filters, thus the features the architecture desires to highlight, increases. A \ac{CNN} usually has three types of layers: (i) convolutional layers, to perform the convolution operations to the input, (ii) pooling layers, to discretize the input and reduce the number of learnable parameters; and (iii) fully connected layers, that are essentially feed-forward neural networks, usually placed at the end of the architecture. The goal of the fully connected layers is to hold the high-level features found during the convolutions and try to learn non-linear combinations of these features before assigning the input image a label. Details about these layers contextualized in our model are provided in Section~\ref{section:cnn_details}.

One of the fundamental decisions to be taken when designing a \ac{CNN}, or generically a neural network, concerns the representation of the input data. Several input representations are available in the literature, each bringing its advantages and drawbacks. Although for visual images the choice is straightforward, for audio samples numerous alternatives are possible, including \ac{MFCC}, raw digitized sample stream, machine discovered features, and hand-crafted features. Even if the best input representation to adopt is strongly dependant on the problem to solve, several studies in the literature show that feeding \ac{CNN} with spectrograms is effective in many fields, including musical onset detection~\cite{schluter}, human detection and activity classification~\cite{kim}, music classification~\cite{costa}, and other interesting activities~\cite{Wyse2017AudioSR}.

\subsection{Guns and Gunshots}

\begin{figure}[htbp]
    \centering
    \includegraphics[width=\columnwidth]{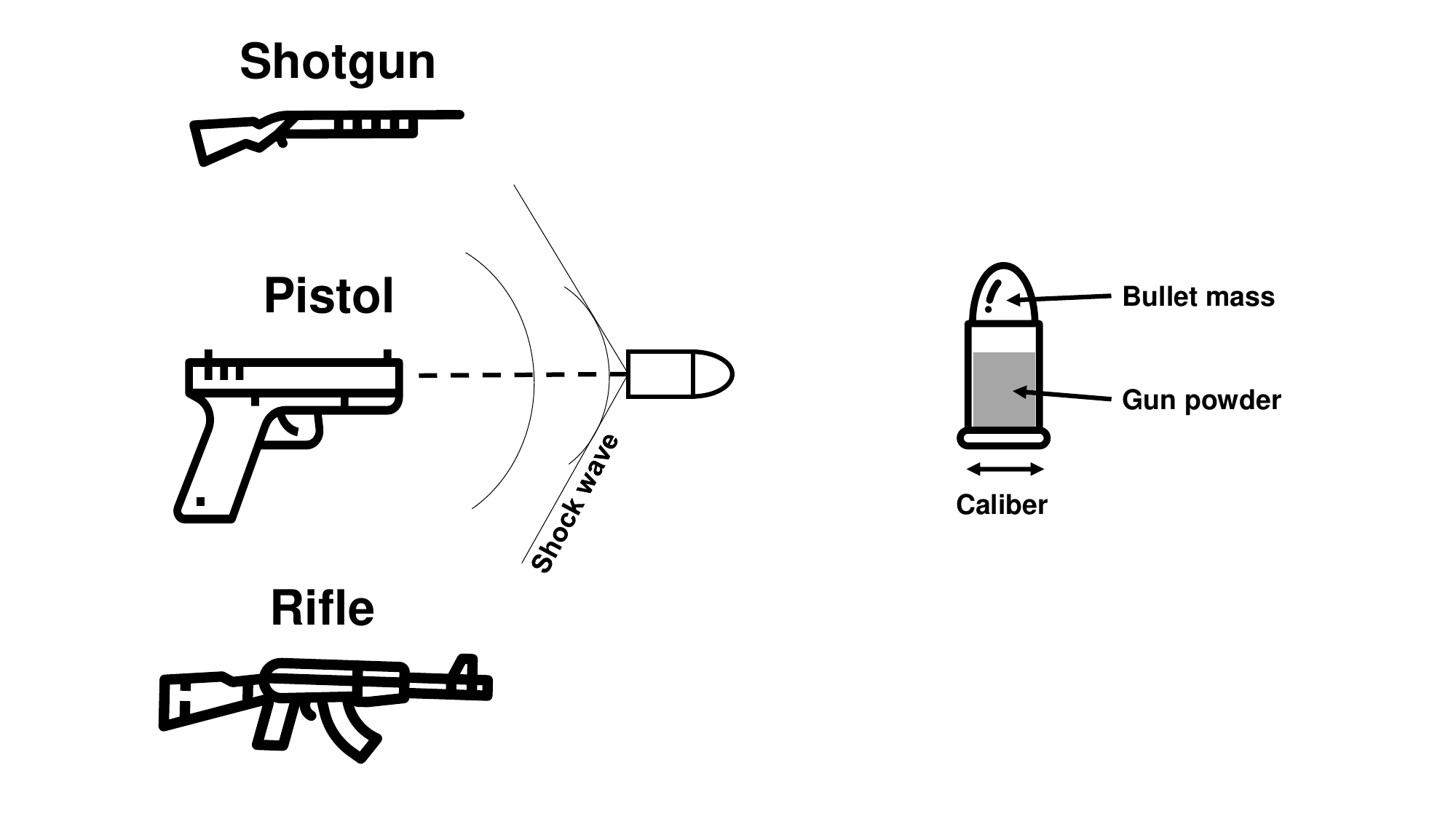}
    \caption{Variables taken into account in our analysis: category of firearm, caliber, and gun model.}
    \label{fig:guns_and_shot}
\end{figure}

Gunshots are the result of multiple acoustic events, namely, the \emph{muzzle blast} created by the explosion inside the barrel and the \emph{ballistic shockwave} that is generated by the supersonic projectile. These phenomena are the results of many characteristics and variables that eventually sum up and generate the acoustic blast, which include the firearm type, model, barrel length, ammunition type, powder quantity, weight and shape of the projectile, and possibly others. The aim of this work is to estimate at what extent it is possible to use a gunshot as a unique fingerprint that uniquely identifies one or more of the aforementioned variables. Figure~\ref{fig:guns_and_shot} summarizes the most important characteristics affecting the acoustic blast generated by a gun.

Our observation is that different configurations of the aforementioned parameters may lead to unique gunshot patterns that can be detected by analyzing the frequency-time decomposition of the gunshot blast. In the next sections, we demonstrate how Convolutional Neural Networks (CNNs) can be effectively used to detect these patterns, thus uniquely identifying the category of gun, the caliber, and finally, the model of the gun.

\section{Dataset description}
\label{sec:dataset_description}
Table~\ref{table:dataset} shows the dataset considered in this work. We collected the samples from several YouTube videos, such as \emph{C4Defense}, \emph{hickok45}, \emph{EmanuelRJSniper}, \emph{mixup98}, \emph{OneGear}, and \emph{ReloaderJoe}. Our choice of guns takes into account two main aspects: the \emph{Category of Guns} and the \emph{Caliber}.

{\bf Category of guns.} We considered 30 different \emph{pistols}, 18 \emph{rifles}, and 11 \emph{shotguns}. As for pistols, we considered 22 \emph{revolvers} and 10 \emph{semiautomatic}. 

{\bf Caliber.} We took into account the most popular calibers in U.S. and world-wide~\cite{calibers,calibers2}, such as 9mm and .45acp for automatic pistols, .44M and .357M for revolvers, 7.62x39 and 5.56NATO for rifles, and 12 gauge caliber for shotguns.

\begin{table}
\centering
\caption{Dataset: Gun Model, Caliber, and number of extracted samples.}
\label{table:dataset}
\begin{adjustbox}{max width=0.95\columnwidth}
\begin{tabular}{|l|c|c|r|r|}
\hline
{\bf Gun Model} & {\bf Caliber} & {\bf Category} & {\bf N. Samples} & {\bf ID} \\ \hline\hline
Glock 45 & 9mm & Pistol & 154 & 24 \\ \hline
Beretta 98 FS & 9mm & Pistol & 29  & 2 \\ \hline
Beretta 92 FS & 9mm & Pistol &  361 & 11 \\ \hline
Beretta PX4 Storm & 9mm & Pistol & 164 & 12 \\ \hline\hline
Glock 21 & .45acp & Pistol & 145 & 22 \\ \hline
M\&P Shield & .45acp & Pistol &  77 & 27 \\ \hline
Colt 1911 & .45acp & Pistol &  81 & 34 \\ \hline
Walther PPQ & .45acp & Pistol & 115 & 57 \\ \hline
Glock 30S & .45acp & Pistol & 103 & 23 \\ \hline\hline
S\&W 629 4-inch & .44M & Pistol  & 97 & 5 \\ \hline
S\&W 629 TrailBoss & .44M & Pistol  & 62 & 6 \\ \hline
S\&W 629 Performance Center & .44M & Pistol & 42 & 7 \\ \hline
S\&W 629-8 & .44M & Pistol  & 48 & 47 \\ \hline
S\&W 69 & .44M & Pistol  & 35 & 49 \\ \hline
S\&W 69 2.75-inch & .44M & Pistol  & 35 & 50 \\ \hline
Charter Arms .44 Special Bulldog & .44M & Pistol  & 40 & 15 \\ \hline
S\&W Model 29 Dirty Harry & .44M & Pistol  & 42 & 21 \\ \hline      
S\&W Model 29 4-inch & .44M & Pistol  & 51 & 32 \\ \hline 
Ruger Redhawk Big Game Hunt & .44M & Pistol  & 23 & 42 \\ \hline 
Ruger Super Black Hawk & .44M & Pistol  & 25 & 44 \\ \hline\hline
Ruger Red Hawk 8-shots & .357M & Pistol & 48 & 41 \\ \hline
S\&W 357Magnum & .357M & Pistol  & 73 & 4 \\ \hline
Chiappa Rhino & .357M & Pistol & 47 & 16 \\ \hline
Coonan 1911 & .357M & Pistol & 56 & 17 \\ \hline
Ruger GP100 Match Champion & .357M & Pistol & 36 & 38 \\ \hline
Ruger SP101 & .357M & Pistol & 45 & 43 \\ \hline
S\&W Model 19 3-inch & .357M & Pistol & 42 & 45 \\ \hline 
S\&W Model 27 & .357M & Pistol & 42 & 46 \\ \hline 
S\&W Model 66 & .357M & Pistol & 54 & 48 \\ \hline 
Dan Wesson Revolver & .357M & Pistol & 30 & 19 \\ \hline\hline 
CZ Bren 2 MS & 7.62x39 & Rifle & 49 & 1 \\ \hline
PWS MK107 & 7.62x39 & Rifle & 73 & 3 \\ \hline
CZ 527 & 7.62x39 & Rifle & 38 & 13 \\ \hline 
Century Arms C39 AK-47 & 7.62x39 & Rifle & 37 & 14 \\ \hline
Maadi AK47 & 7.62x39 & Rifle & 60 & 28 \\ \hline    
Micro Draco AK47 Pistol & 7.62x39 & Rifle & 44 & 29 \\ \hline    
N-PAP AK & 7.62x39 & Rifle & 46 & 33 \\ \hline  
Ruger American Ranch Rifle & 7.62x39 & Rifle & 38 & 37 \\ \hline  
Ruger Mini 30 & 7.62x39 & Rifle & 51 & 40 \\ \hline  
SKS & 7.62x39 & Rifle & 26 & 59 \\ \hline  
Daniel Defense M4 A1 SOCOM & 5.56 NATO & Rifle & 68 & 20 \\ \hline
Ruger AR & 5.56 NATO & Rifle & 78 & 36 \\ \hline
Ruger Mini-14 & 5.56 NATO & Rifle & 60 & 39 \\ \hline 
Ruger AR556 MPR & 5.56 NATO & Rifle & 30 & 35 \\ \hline 
SIG 556 Classic SWAT Model & 5.56 NATO & Rifle & 72 & 51 \\ \hline 
Springfield Armory Saint & 5.56 NATO & Rifle & 68 & 54 \\ \hline
Tactical Edge Warfighter & 5.56 NATO & Rifle & 49 & 56 \\ \hline 
M\&P 15 Sport II & 5.56 NATO & Rifle & 84 & 26 \\ \hline\hline
Benelli M2 SBS & 12 & Shotgun & 44 & 8 \\ \hline
Benelli M4 & 12 & Shotgun & 45 & 9 \\ \hline
Benelli Nova & 12 & Shotgun & 36 & 10 \\ \hline
DP-12 & 12 & Shotgun & 74 & 18 \\ \hline
Kel-Tec SG12 & 12 & Shotgun & 52 & 25 \\ \hline
Winchester Model 12 & 12 & Shotgun & 46 & 31 \\ \hline
SRM 1216 & 12 & Shotgun & 34 & 52 \\ \hline
Serbu Super Shorty & 12 & Shotgun & 33 & 53 \\ \hline
Standard Manufacturing SKO Shorty & 12 & Shotgun & 47 & 55 \\ \hline
Winchester SXP Defender & 12 & Shotgun & 36 & 58 \\ \hline
Winchester Model 12 SlugFest & 12 & Shotgun & 35 & 30 \\ \hline
\end{tabular}
\end{adjustbox}
\end{table}

\subsection{Muzzle blast: preliminary considerations}
\label{sec:muzzle_blast}
When a gun is fired, there are two distinct acoustic phenomena, the \emph{muzzle blast} and the \emph{ballistic shockwave}~\cite{sallai}. The latter is generated by the bullet that compresses the air in front of itself creating a sonic boom that propagates with a shape of a cone where the vertex is the bullet itself. Conversely, the muzzle blast is a high energy acoustic signal originated by the gun's muzzle with a spherical wavefront, propagating at the speed of sound, and with center the muzzle of the gun.
The ballistic shockwave is a very important source of information to locate a sniper in an open field~\cite{sallai2, volgyesi}. However, to achieve that, the ballistic shockwave has to be sampled from different locations requiring an array of microphones. The ballistic shockwave cannot be observed for subsonic projectiles such as those used in shotguns and pistols. 

Given the aforementioned considerations, we focus on the muzzle blast and the echoes associated with it. In the following, we discuss and highlight three critical parameters that have to be carefully set in order to maximize the detection performance of a neural network: (i) \emph{muzzle blast duration}, (ii) \emph{number of frequency bins}, and (iii) the \emph{number of time slots}.

\begin{figure}[htbp]
    \includegraphics[width=\columnwidth]{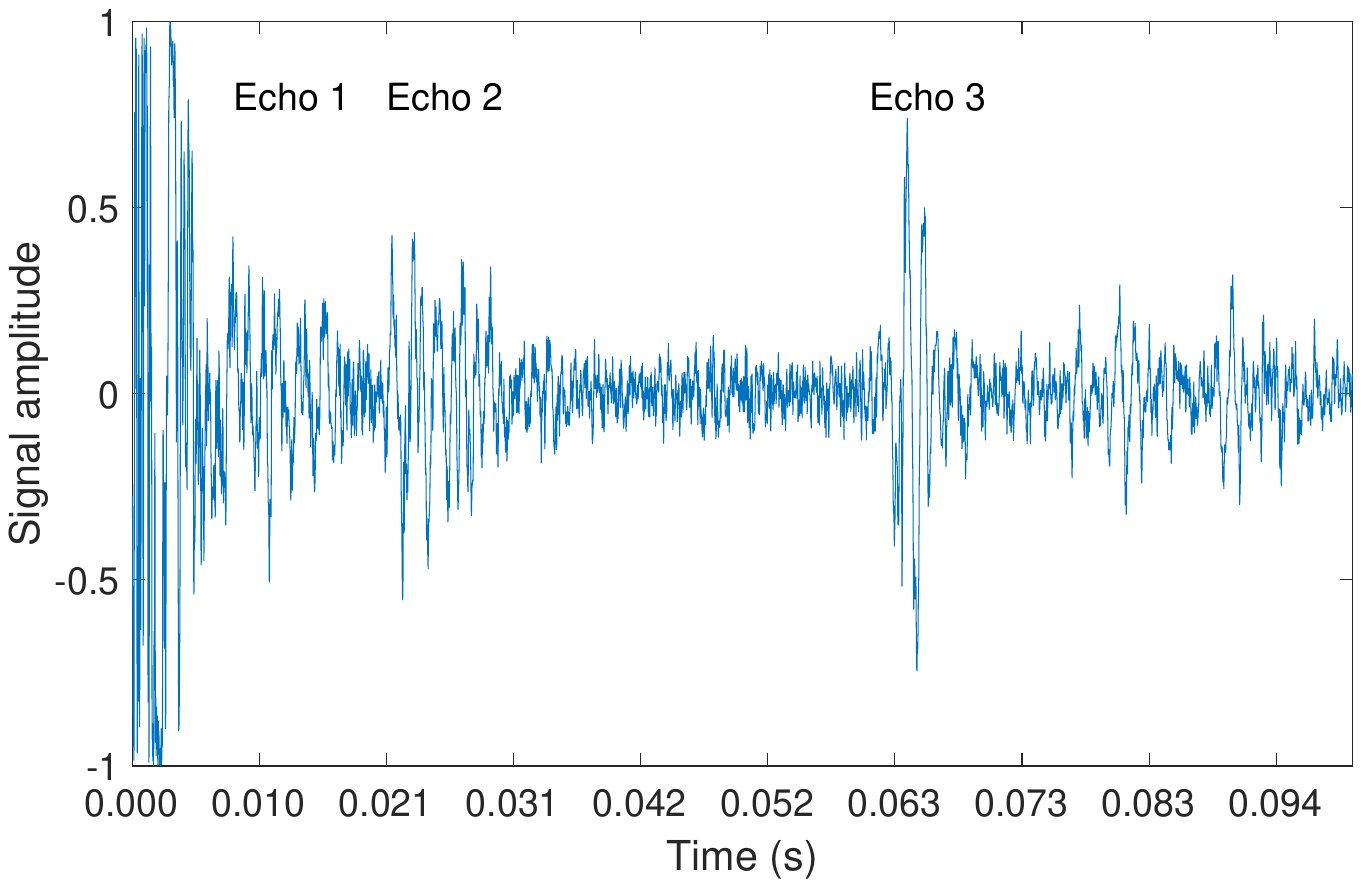}
    \centering
    \caption{Acoustic signal amplitude of a muzzle blast for a Beretta PX4 Storm, 9mm.}
    \label{fig:sound_sample}
\end{figure}

Figure~\ref{fig:sound_sample} shows the acoustic signal amplitude recorded from a Beretta PX4 Storm, 9mm. The muzzle blast lasts for a few milliseconds (up to 5ms in the figure), depending on the model of gun and caliber. We also observe some echo effects (Echo 1, Echo 2, and Echo 3) at 10ms, 22ms, and 63ms due to reflections of the sound from obstacles around the shooter. We highlight that this is consistent with previous findings from other studies~\cite{sallai}, while the \emph{muzzle blast duration} will be a critical parameter from the analysis carried out in this work.

\begin{figure}[!htbp]
    \centering
    \includegraphics[width=\columnwidth,height=60mm]{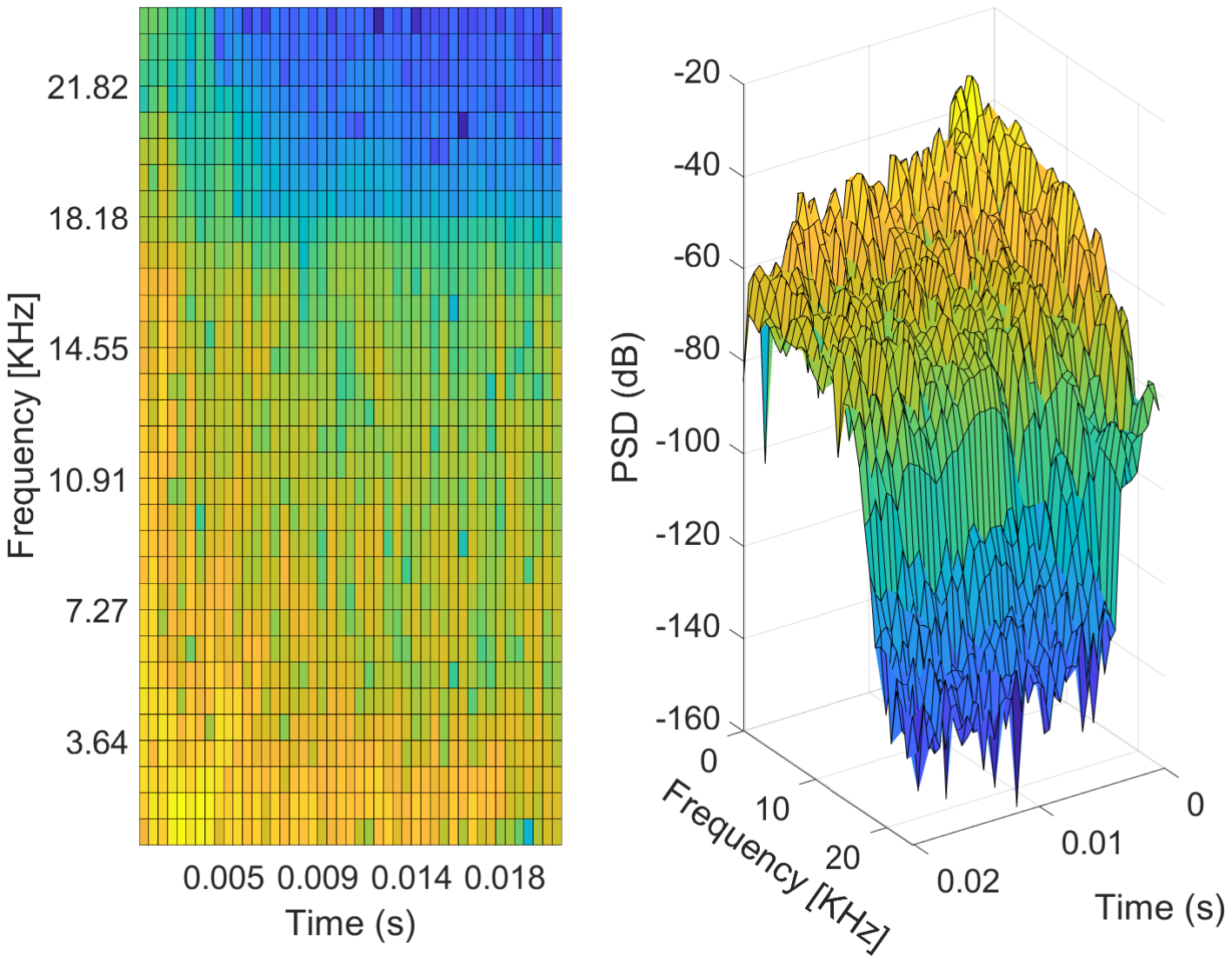}
    \caption{Spectrogram of a muzzle blast: bi-dimensional and three-dimensional \ac{PSD} of a muzzle blast (Beretta PX4 Storm, 9mm) as a function of time and frequency.}
    \label{fig:sound_sample_freq} 
\end{figure}

Figure~\ref{fig:sound_sample_freq} shows the \ac{PSD} as a function of time and frequency (spectrogram) associated with the muzzle blast in Fig.~\ref{fig:sound_sample}. We consider both the bi-dimensional and the three-dimensional representation of the spectrogram. We observe that the muzzle blast (time less than 5ms) takes all the frequency components between 0 and 24KHz with a significant power spanning between -30dB (lower frequencies) and -80dB (higher frequencies). As soon as the blast finishes, the echoes take the frequencies less than 18KHz with a decreasing power between -40dB and -60dB. The aforementioned spectrogram components constitute the input for the training process of our neural network. 

We identify two more critical parameters affecting our algorithm performance: the \emph{number of frequency bins} and the \emph{number of time slots}. For our analysis, we adopted the \emph{spectrogram} function of MATLAB-R2019b considering as input the acoustic sample (0.1 seconds from the beginning of the blast), a window of size $w=44$ to divide the signal into segments and performing windowing according to the Hann function, $no = \lfloor{w/2}\rfloor$ as the number of overlapping samples between adjacent segments, $fl = 65$ as the FFT length, and $fs = 48000$ as the number of samples per seconds acquired by the microphone. Assuming the previous parameters, the \emph{spectrogram} function returns the \ac{PSD} of $(fl + 1)/2$ frequencies and $\lfloor{\frac{length(x) - no}{w - no}}\rfloor$ time bins, where $x$ is the vector of the acoustic samples being equal to $0.21 \cdot fs =10080$ sample---we considered the first 0.21 seconds after the first abrupt change as per Fig.~\ref{fig:sound_sample}. For instance, in the previous example, the frequency range (0 to 24KHz) has been divided into 33 bins, while the time has been divided into 46 slots.    

\subsection{Quality of the audio samples}
\label{sec:quality_of_the_audio_samples}

\begin{figure}[htbp]
    \includegraphics[width=\columnwidth]{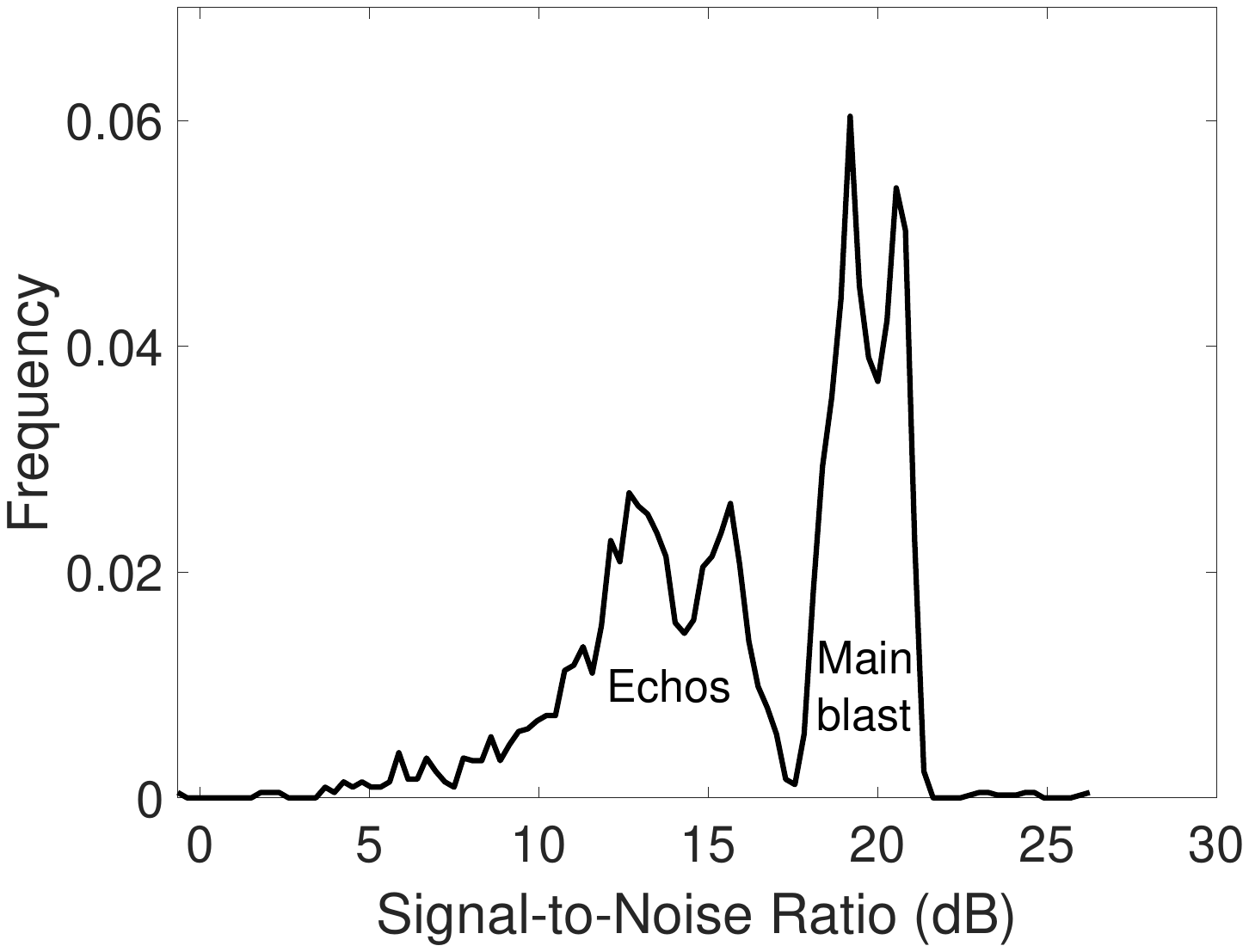}
    \centering
    \caption{Sound quality analysis: \ac{SNR} of a main blast and its associated echos assuming a reference random noise of amplitude one tenth of the microphone saturation threshold.}
    \label{fig:sound_quality}
\end{figure}

In the following, we provide a quantitative analysis of the quality of the collected audio samples. As a quality metric, we consider the \ac{SNR} computed on each muzzle blast from the actual starting of the blast for a period of 400ms. For each audio sample, we consider a pre-defined reference noise pattern constituted by random samples of amplitude 0.1, i.e., one-tenth of the maximum signal amplitude taken by the microphone. The previous sound pressure is equivalent to a classical background noise that can be sampled from an outdoor environment characterized by a gentle wind. Figure~\ref{fig:sound_quality} shows the probability distribution function associated with the \ac{SNR} computed as described before. The overall audio quality is very high since the muzzle blast is +20dB higher than the reference noise pattern. We observe that even the echoes can be easily identified from the noise reference.

\section{Dataset generation}
\label{sec:dataset_generation}

We generated a dataset of 3655 samples extracted from videos found on YouTube. Each of the collected audio samples has a sample rate of either 48000 or 44100 samples per second. Generating a dataset of gunshots extracted from YouTube videos involves the following steps:

\begin{itemize}
    \item {\bf Audio extraction.} We performed the audio extraction (MP3 format) from the selected videos using the \emph{youtube-dl}~\cite{youtube-dl} and \emph{ffmpeg}~\cite{ffmpeg} tools.
    \item {\bf Abrupt change detection.} A preliminary filtering is performed by identifying abrupt changes in the audio signal. 
    \item {\bf Gunshot detection.} Gunshots are detected among blasts by relying on a \ac{SVM} learning algorithm.
\end{itemize}

In the following, we describe the procedure of automatically extracting gunshots from an audio trace focusing on \emph{Blast detection} and \emph{Gunshot detection}.

\subsection{Identification of abrupt changes in an audio trace}
\label{sec:blast_detection}

\begin{figure}[htbp]
    \includegraphics[width=\columnwidth]{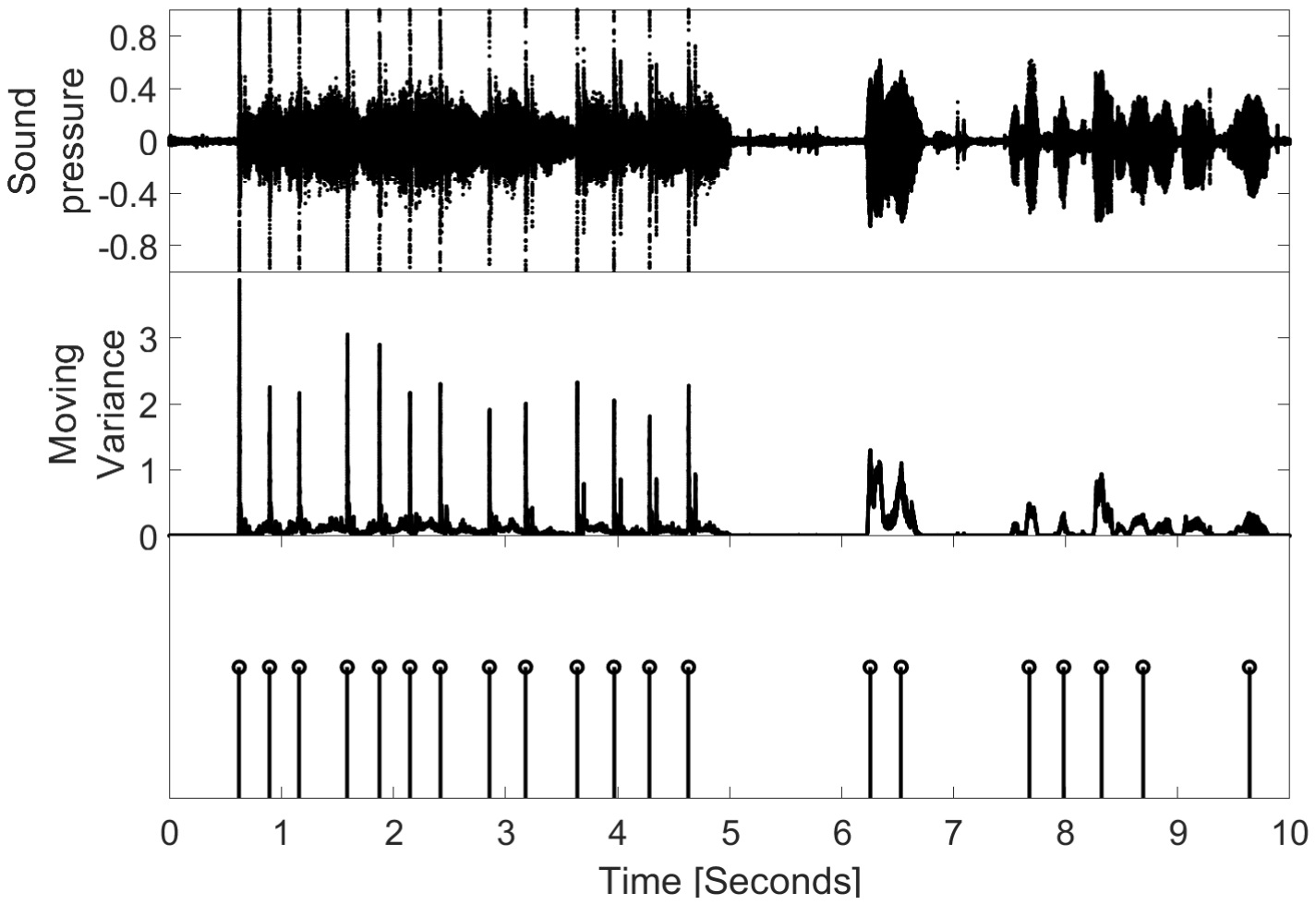}
    \centering
    \caption{Detection of abrupt changes in audio traces: from sound pressure to abrupt change detection by computing moving variance and peak detection.}
    \label{fig:blast_detection}
\end{figure}

To detect \emph{abrupt changes} in an audio trace, we computed the variance over a sliding window of 5ms, equivalent to either 220 or 240 samples depending on the quality of the audio trace, i.e., 44100 or 48000 samples per second, respectively. Subsequently, we searched for the peaks adopting windows of size 0.3 seconds and a minimum peak prominence of 0.3. Fig.~\ref{fig:blast_detection} shows the three computation stages from the sound pressure to the blast sequences that are passing by the moving window averaging. This figure refers to two sound chunks extracted from an audio trace, where the first part (i.e., $0 \le t \le 5.5$ seconds) is a sequence of gunshots, while the second part (i.e., $t > 5$ seconds) is mainly constituted by voice. We stress that the main aim of this part is to detect abrupt changes in the sound pressure, while subsequently we will show how gunshots are identified.

\subsection{Gunshot detection}
\label{sec:shot_detection}

\begin{figure}[htbp]
    \includegraphics[width=\columnwidth]{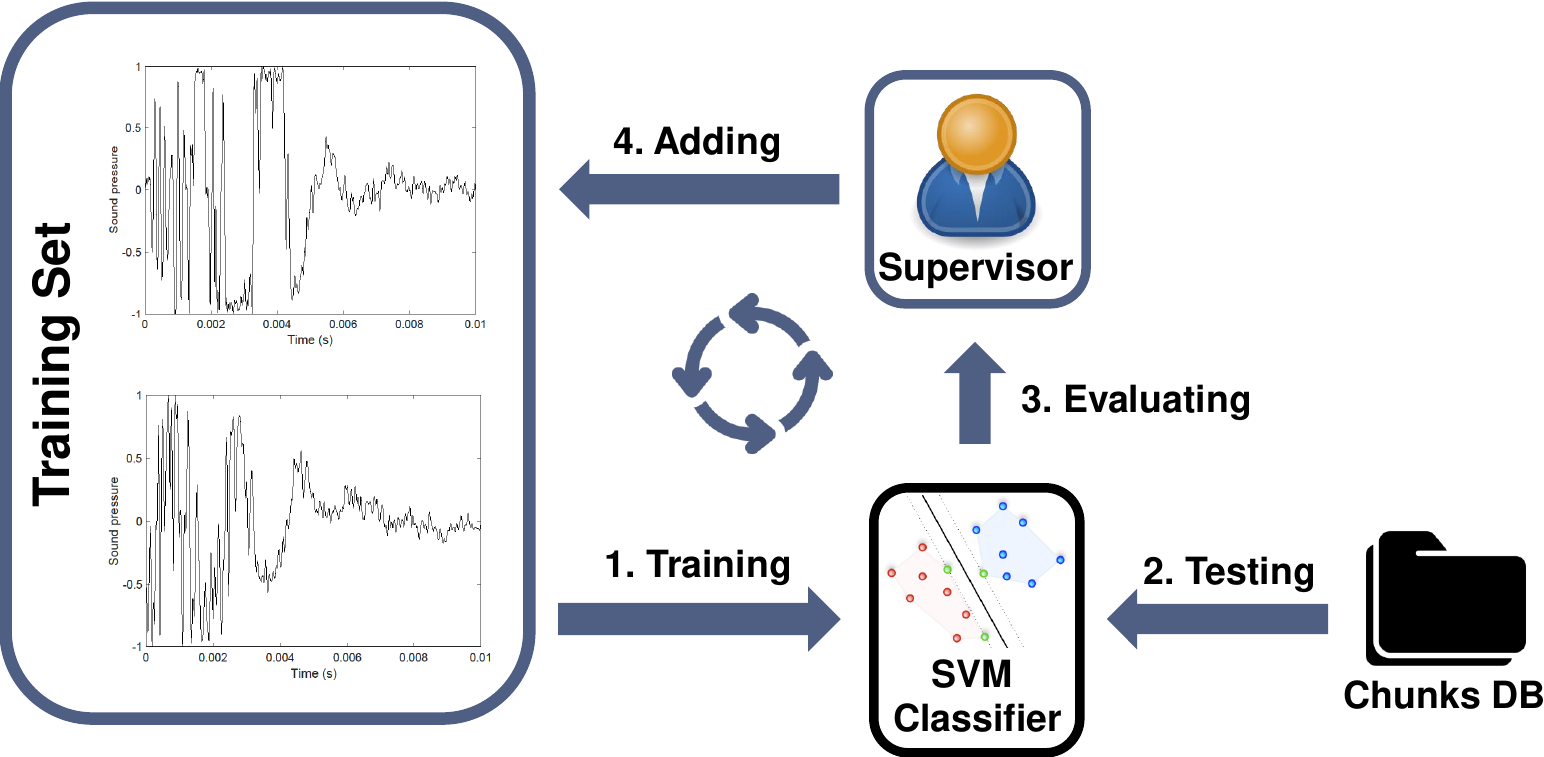} 
    \centering
    \caption{Gunshot detection via human-assisted supervised learning. The \ac{SVM} classifier is trained by verified samples of gunshots. When new gunshot samples are tested, the output of the classifier is verified (by the user) and added to the Training Set.}
    \label{fig:svm_classifier}
\end{figure}

Gunshot detection is performed via a \emph{human-assisted supervised learning} approach. The intention 
is to have a growing training set of actual gunshots that is supervised by the user. The user checks for both false positives and false negatives by listening to the newly generated samples in the training set. 
Figure~\ref{fig:svm_classifier} shows the training, validation, and testing procedures. We assume that the training set is populated with an initial dataset of actual gunshots that have been manually selected. In our case, we started from an initial dataset of 10 gunshot samples only. At each cycle, a new model is trained with the current training set (Step 1 in Fig.~\ref{fig:svm_classifier}). Subsequently (Step 2 in Fig.~\ref{fig:svm_classifier}), new samples are selected from the list that is generated by the procedure presented in Section~\ref{sec:blast_detection}. Finally, the generated samples are tested with the current training set. The output is assessed by the supervisor (Step 3 in Fig.~\ref{fig:svm_classifier}), and the verified samples are added to the training set (Step 4 in Fig.~\ref{fig:svm_classifier}).

\begin{figure}[!htbp]
    \centering
    \includegraphics[width=\columnwidth]{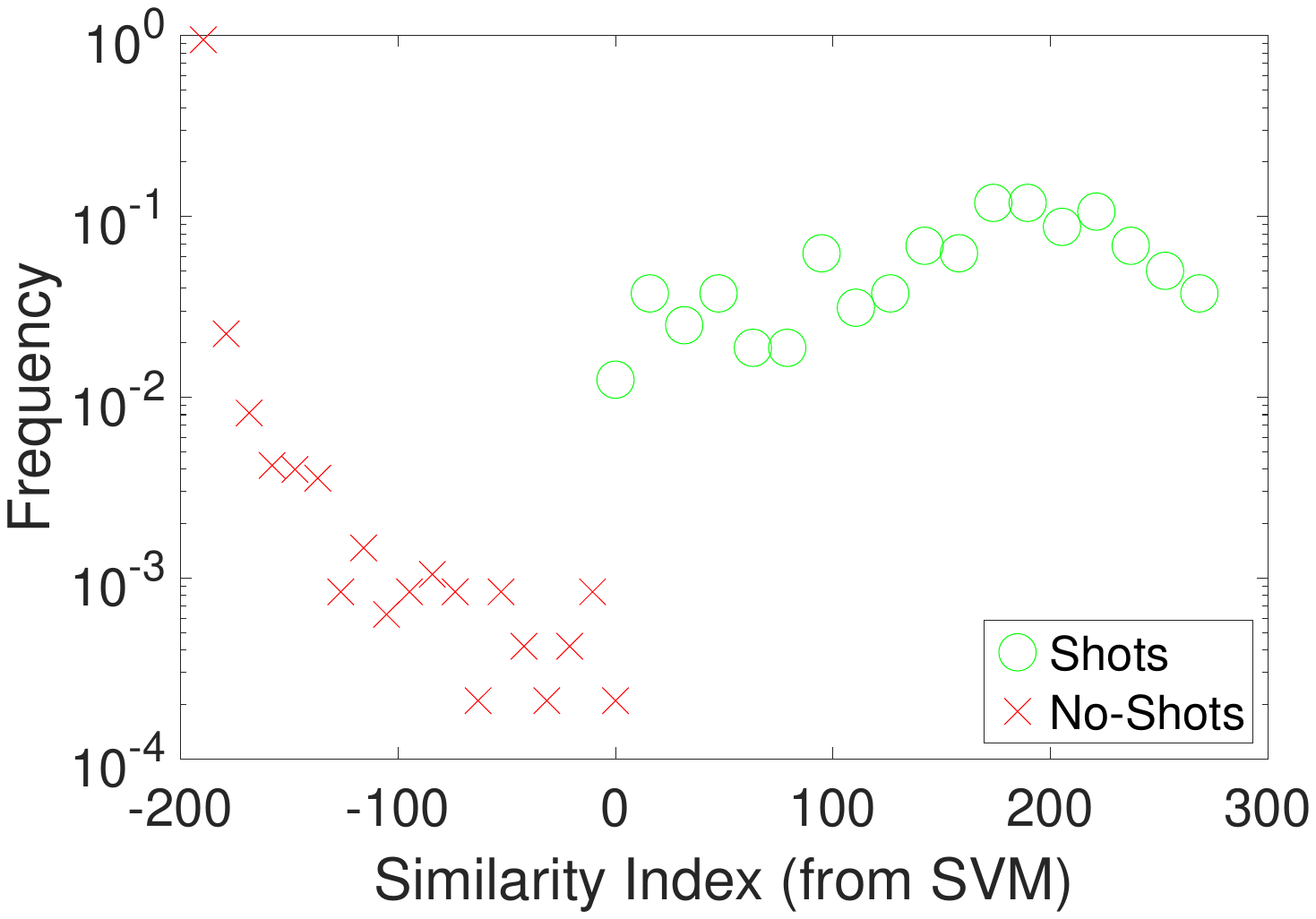}
    \caption{Gunshot detection performance: frequency distribution of Shot and No-Shot samples as a function of their similarity index.} 
    \label{fig:shot_detection}
\end{figure}

{\bf Classification performance.} To assess the quality of the classification procedure, we considered 6 additional videos (V1, \ldots, V6) downloaded from YouTube, which are not included in the training set. For each video, we detected the abrupt changes according to the procedure presented in Section~\ref{sec:blast_detection}, and we executed the gunshot detection procedure presented in Fig.~\ref{fig:svm_classifier}. As for the Training Set, we considered the one we generated from the samples found in Table~\ref{table:dataset}. Figure~\ref{fig:shot_detection} shows the frequency of the similarity indexes provided by the \ac{SVM} classifier for the Shot and No-Shot audio samples with red crosses and green circles, respectively. The similarity indexes were categorized into bin width of 10, where each cross/circle aggregates adjacent similarity indexes. Figure~\ref{fig:shot_detection} represents the decision after one iteration of the procedure presented in Fig.~\ref{fig:svm_classifier}. The decision Shot vs No-Shot is taken as a function of the threshold $Thr$, which has been empirically set to zero. We observed that 96\% of the No-Shot samples feature a similarity index of -189.5, while the remaining 4\% are spread between -178.9 and -0.55. There are no samples from the No-Shot class with a similarity index that is greater than 0. As for the Shot class, the samples are distributed between 0.41 and 275, with frequencies between 1\% and 11\%. Even in this case, we highlight that there are no samples from the Shot class with a similarity index that is less than 0. 

To precisely assess the effectiveness of our solution, we manually checked all of the classified samples, namely, Shot vs No-Shot. Table~\ref{table:svm_performance} shows the result of our analysis. For each video, we report the number of detected abrupt changes ($N$), the threshold used by the \ac{SVM} classifier ($Thr=0$), True Positive ($TP$), False Positive ($FP$), True Negative ($TN$), False Negative ($FN$), the actual number of gunshots ($Actual$), and the overall accuracy of the detection algorithm. As previously stated, during our evaluation, we considered only one iteration as depicted in Fig.~\ref{fig:svm_classifier}. We would like to highlight that the proposed algorithm achieves 
the main purpose of generating a dataset of gunshot samples (i) in a fast and efficient way and (ii) with the minimum amount of false positives. The output of this phase will be the training set to be used by the \ac{CNN}. 

At this stage, we aim at minimizing the number of $FP$, which might bias the subsequent training process. We also aim at maximizing the process efficiency of creating a large dataset of gunshot samples. Therefore, the task of the supervisor  resorts mainly to listening to a very few samples ($TP+FP$) despite the dataset $N$, in order to remove the $FP$, which are overall very few: only 2 out of 4931 samples. Conversely, we observe that our approach might lose some good samples ($FN = 16+4$). 
However, these samples do not affect the performance of our solution hence we consider them not important.

The above procedure has been applied to each audio sample found in Table~\ref{table:dataset} to generate a dataset of actual gunshot samples, that is, one dataset for each gun model.

\begin{table}[]
\caption{Shot detection performance considering 6 videos}
\centering
\begin{adjustbox}{max width=0.95\columnwidth}
    \begin{tabular}{r|r|r|r|r|r|r|r|c|}
    \cline{2-9}
    & \multicolumn{1}{c|}{\textbf{N}} & \multicolumn{1}{c|}{\textbf{Thr}} & \textbf{TP} & \textbf{FP} & \textbf{TN} & \textbf{FN} & \textbf{Actual} & \textbf{Accuracy} \\ \cline{2-9} 
    \textbf{V1} & 643  & 0 & 24 & 2 & 617  & 0  & 24 & {\bf 0.99} \\
    \textbf{V2} & 385  & 0 & 4  & 0 & 365  & 16 & 20 & {\bf 0.96} \\
    \textbf{V3} & 79   & 0 & 5  & 0 & 74   & 0  & 5  & {\bf 1} \\
    \textbf{V4} & 32   & 0 & 5  & 0 & 27   & 0  & 5  & {\bf 1} \\
    \textbf{V5} & 1860 & 0 & 88 & 0 & 1772 & 0  & 88 & {\bf 1} \\
    \textbf{V6} & 1932 & 0 & 31 & 0 & 1897 & 4  & 35 & {\bf 0.99} \\ \cline{2-9} 
    \end{tabular}
\end{adjustbox}
\label{table:svm_performance}
\end{table} 

\section{Overall Architecture}
\label{sec:overall_architecture}

\begin{figure*}[!htbp]
    \centering
    \begin{adjustbox}{width=0.9\textwidth}
    \includegraphics{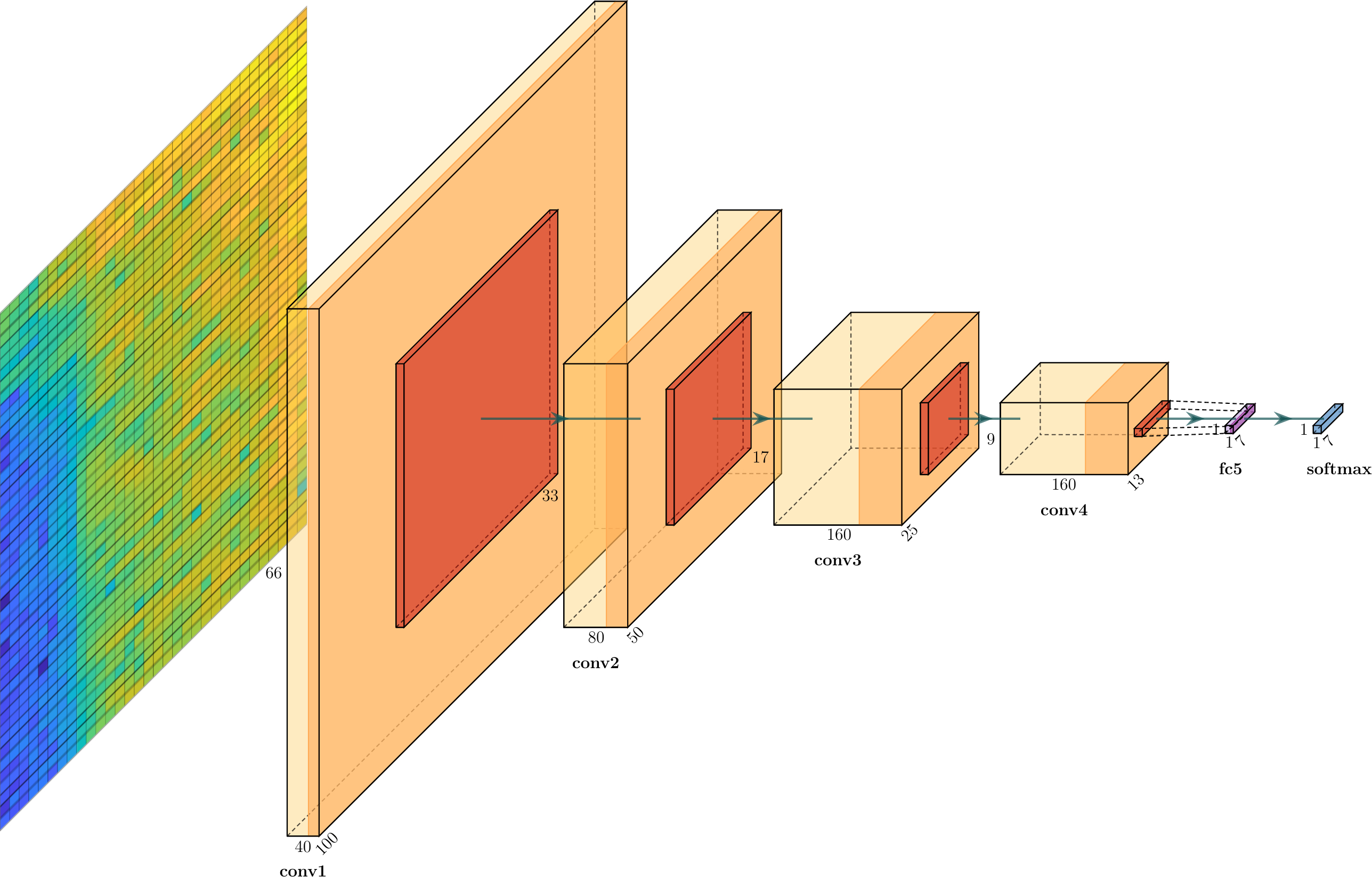}
    \end{adjustbox}
    \caption{Structure and details of our \acl{CNN}.}
    \label{fig:cnn}
\end{figure*}

Figure~\ref{fig:cnn} depicts the overall architecture of our \ac{CNN} consisting of five layers with weights: the first four are convolutional layers, while the last one is a fully connected layer. The output of the fully connected layer is fed to a 7-way softmax, that outputs the probability distribution over the 7 class labels. The details of our architecture, including information about the layers and their learnable parameters, are reported in Appendix~\ref{table:makingup-layers}.

Considering the dimension of the starting image and the need to give importance also to peripheral pixels, every convolution of our architecture makes use of padding to avoid losing information. By adding additional pixels to the border, every convolution layer outputs an image with the same number of pixels as the one fed into that layer. Furthermore, in our \ac{CNN} architecture, we make use of a stride of 1 during convolutions, and a stride of 2 during the Max Pooling application. The stride is a critical hyperparameter in the context of \ac{CNN}, as it allows us to specify the number of cells by which filters (e.g., convolution filters, pooling filters) slides over the image. If the stride is equal to 2, the filter starts from the top left corner and moves over the image with jumps of 2 units at a time. By considering square filters (i.e., \textit{f}x\textit{f}) and square initial images (i.e., \textit{n}x\textit{n}), after having specified the dimension of the filters \textit{f}, whether they are convolutional filters or pooling filters, the stride parameter \textit{s}, the dimensions of the initial images \textit{n}, and the padding \textit{p}, it is possible to calculate the dimension of the square output image of a layer as:

\begin{equation}\label{equation:output_dimension}
    \left\lfloor\frac{n + 2p - f}{s} + 1\right\rfloor 
\end{equation}

Our choice to keep a unit stride during the convolutions and a stride equals to 2 during the pooling is guided by the intention of not losing information during convolution phases, while exploiting the pooling technique to summarize the features, thus reducing the input dimensionality. 

The first convolutional layer filters the 36 x 99 x 1 spectrogram image with 40 kernels of size 3 x 3 x 1, without any stride and with 'same' padding. Setting the padding to ``same'' allows the classifier to calculate and add the padding so that the output has the same size as the input. The second convolutional layer takes as input the normalized (40 channels) and pooled (3x3 max pooling, stride = 2) output of the first convolutional layer and filters it with 80 kernels of size 3 x 3 x 40. The third convolutional layer takes as input the normalized (80 channels) and pooled (3x3 max pooling, stride = 2) output of the second convolutional layer and filters it with 160 kernels of size 3 x 3 x 80. The fourth convolutional layer takes as input the normalized (160 channels) and pooled (3x3 max pooling, stride = 2) output of the third convolutional layer and filters it with additional 160 kernels of size 3 x 3 x 160. The normalized (160 channels) and pooled (1x13 max pooling, stride = 2) output of the fourth convolutional layer is fed to a 7-neuron fully connected layer that, in turn, outputs the result to a 7-way softmax, that produces a distribution over the 7 class labels.

\subsection{\ac{CNN} Details} \label{section:cnn_details}

\textbf{Activation Function}. Our neural network relies on the \ac{ReLU} activation function~\cite{nair} after each convolution. The \ac{ReLU} activation function, whose operation is simplified in equation~\ref{equation:ReLU}, outputs the maximum value between zero and the input value.

\begin{equation}\label{equation:ReLU}
f(x) =
    \begin{cases} 
      x & \text{if $x \geq 0$}, \\
      0 & \text{otherwise}
    \end{cases}
\end{equation}

Although the literature 
uses other variants (e.g., Tanh, SoftSign, Sigmoid), several studies show that \ac{ReLU} outperforms the competitors in terms of performance~\cite{glorot, krizhevsky}. \\
\textbf{Regularization.} Our neural network relies on Dropout~\cite{srivastava} regularization to reduce the likelihood of overfitting. The Dropout regularization technique allows the neural network to randomly cut out units (together with their connections) during the training phase with a given probability. This discourages neurons to rely on the presence of particular other neurons and forces them to find more robust features with different ones~\cite{krizhevsky}, thus reducing the probability of learning the training set by heart. \\
\textbf{Normalization.} Training neural network without normalization leads to the internal covariate shift phenomena, where the distribution of each layer's input change during training, thus requiring a more sophisticated tuning of the parameters. To mitigate this issue we add Batch Normalization layers after each convolution. The Batch Normalization technique~\cite{bach} performs the normalization for each training mini-batch, allowing the usage of higher learning rates and reducing the need for a cherry-picking tuning of the parameters. As summarized in the equation~\ref{equation:normalization}, Batch Normalization normalizes the output of an activation layer by subtracting the mean and dividing by the standard deviation of the batch. \\ 
Given a mini-batch $\beta = {x_1, ..., x_m}$: 

\begin{equation}\label{equation:minibatch-mean}
\mu_\beta \leftarrow \frac{1}{m} \sum_{i=1}^{m} x_i
\end{equation}

\begin{equation}\label{equation:minibatch-variance}
\sigma^2_\beta \leftarrow \frac{1}{m} \sum_{i=1}^{m} (x_i - \mu_\beta)^2
\end{equation}

\begin{equation}\label{equation:normalization}
\hat{x} \leftarrow \frac{x_i - \mu_\beta}{\sqrt{\sigma^2_\beta + \epsilon}}
\end{equation}

where $\epsilon$ is defined as a constant to add to the mini-batch variances, specified as a scalar $\ge$ $10^{-5}$.

Although the Batch Normalization technique brings a slight regularization effect to the neural network, in some cases eliminating the need for Dropout~\cite{bach}, we find that the combined use of the Batch Normalization and Dropout aids generalization~\cite{krizhevsky}. \\
\textbf{Discretization.} The application of discretization techniques to an input representation consists of reducing its dimensionality to evaluate the features within the obtained, summarized sub-regions. This process allows us to mitigate the overfitting of the training set and to reduce the number of parameters to be learned for the training, thus reducing the overall computational cost. To attain these benefits, in our architecture we use a Max Pooling sample-based discretization process layer after each activation layer. Max Pooling applies a max filter to non-overlapping sub-regions of the input feature map, whose dimension is dictated by the dimension of the filter. When Max Pooling is applied, the passage of the moving filter onto a sub-region produces, as output, a value, consisting of the maximum value of that sub-region. \\
\textbf{Output.} As for the output layer, our neural network architecture relies on the commonly used softmax function. The softmax function, taking as input a vector of real numbers, produces a probability distribution proportional to the exponential of the input numbers. In detail, the input real numbers are mapped in a (0,1) interval that sums up to one, thus allowing to treat the output provided by the softmax function as probabilities. \\
In general, given a vector of real numbers $v = (v_1, \dots, v_K) \in {\rm I\!R^K}$, the standard unit softmax function $\sigma: {\rm I\!R^K} \rightarrow {\rm I\!R^K}$ is defined by:

\begin{equation}\label{equation:softmax}
    \sigma(v)_i = \frac{e^{v_i}}{\sum_{j=1}^{K} e^{v_j}}
\end{equation}

\subsection{Learning Details}

\begin{table}[h]
    \caption{Training Options of our network.}
    \label{table:training_options}
    \centering
    \begin{tabular}{|c|c|}
        \hline
            \textbf{Option} & \textbf{Value} \\ \hline
            Optimizer & Adam \\ \hline
            InitialLearnRate & $x * 10^{-4}, x \in [1,3]$ \\ \hline
            MaxEpochs & 50 \\ \hline
            MiniBatchSize & 8 \\ \hline
            Shuffle & Every Epoch \\ \hline
            Plots & Training Progress \\ \hline
            Validation Data & Random 20\% of the data \\ \hline
            Validation Frequency & $\lfloor |$training\_set$|$ / miniBatchSize$\rfloor$ \\ \hline
    \end{tabular}
\end{table}

Table~\ref{table:training_options} summarizes the training options of our network, that are detailed in the following. \\
\textbf{Optimizer}. An optimizer is defined as an algorithm (or a method) used to tune the parameters of a neural network with the goal of reducing the loss function. In our architecture, we rely on the \ac{Adam} optimizer~\cite{kingma}, an extensively adopted optimizer that inherits the advantages of both \ac{RMSProp} and \ac{SGD} with momentum (i.e., \ac{SGD} where each gradient update is a linear combination of the previous gradient updates) optimizers.
From \ac{RMSProp} it inherits the squared gradients to scale the learning rate, while from \ac{SGD} with momentum it inherits the concept of the moving average of the gradients. An empirical analysis conducted in~\cite{kingma} shows that Adam outperforms the other optimizers, thus working better in practice. As recommended in the original paper (whose algorithm is reported below with our parameters), in our implementation we set to 0.9 the gradient decay factor $\beta_1$, to 0.999 the squared gradient decay factor $\beta_2$, and to $10^{-8}$ the denominator offset (to avoid divisions by zero), respectively. However, although the original paper recommends using an initial learning rate of $10^{-3}$, we empirically found (relying on the grid search hyperparameter tuning technique) that setting this value to $x * 10^{-4}, x \in [1,3]$ provides better results.

\begin{algorithm}[h]
	\caption{Adam Optimizer~\cite{kingma}}
	\label{algo:adam}
	\textbf{Require:} $\alpha$: Stepsize \\
	\textbf{Require:} $f(\theta)$: Stochastic objective function \\
	\textbf{Require:} $\theta_0$: Initial parameter vector \\
	$m_0 \leftarrow 0$ (Initialization of the $1^{st}$ moment vector) \\
	$v_0 \leftarrow 0$ (Initialization of the $2^{nd}$ moment vector) \\
	$\beta_1 \leftarrow 0.9$ (Initialization of the gradient decay factor) \\
	$\beta_2 \leftarrow 0.999$ (Initialization of the squared gradient decay factor) \\
	$t \leftarrow 0$ (Initialization of the timestep) \\
	$\epsilon \leftarrow 10^{-8}$ (Initialization of the denominator offset)\\
	$\alpha \leftarrow x \cdot 10^{-4}, x \in [1,3]$ (Initialization of the learning rate) \\
	\While{$\theta$ is not converged} {
    	$t = t + 1$ (Increment the timestep) \\
	    $g_t \leftarrow \nabla_\theta f_t(\theta_{t-1})$ (Gradients w.r.t. stochastic objective at timestep t)\\
	    $m_t \leftarrow \beta_1 \cdot m_{t-1} + (1 - \beta_1) \cdot g_t$ (Update biased first moment estimate) \\
	    $v_t \leftarrow \beta_2 \cdot v_{t-1} + (1 - \beta_2) \cdot {g_t}^2$ (Update biased second raw moment estimate) \\
	    $\hat{m_t} \leftarrow \frac{m_t}{1-\beta_1^t}$ (Compute bias-corrected first moment estimate) \\
	    $\hat{v_t} \leftarrow \frac{v_t}{1-\beta_2^t}$ (Compute bias-corrected second raw moment estimate) \\
	    $\theta_t \leftarrow \theta_{t-1} - \alpha \cdot  \frac{\hat{m_t}}{\sqrt{\hat{v_t} + \epsilon}}$ (Update Parameters)
	}
	\Return $\theta_t$ (Parameters)
\end{algorithm}

\noindent \textbf{Number of Epochs}. An epoch is defined as a single pass through the training set, i.e., 1 forward pass and 1 backward pass for all the training samples. A forward pass is defined as the calculation process to obtain the output values from inputs data, from the first layer to the last layers, while a backward pass is defined as the process of changing the weights (i.e., learning) by relying on an optimization algorithm (e.g., the gradient descent algorithm) from the last layer backward to the first layer. We empirically set as 50 the max number of epochs, since each of the subsequent epochs does not bring any benefit to our model learning. \\
\textbf{Mini-Batch Size.} Using mini-batch that consists of processing small subsets of training samples in every iteration, instead of processing them all together. The choice of mini-batch size (e.g., the number of training samples to process) does not affect the performance of the model in terms of accuracy, but affects the resource required during the training process. A larger mini-batch size requires more memory and takes more time per epoch, but allows the classifier to better optimize the vectorization (i.e., the linear transformation of a matrix into a column vector), while a smaller mini-batch size requires less memory but loses the speed-up given by vectorization. In our model, we set the mini-batch size to 8, to better optimize the resources of our server. \\
\textbf{Shuffle.} The ``shuffle'' option allows shuffling the order of which training samples are fed to the model, with the goal of reducing variance, thus reducing overfitting. Shuffling the training samples becomes crucial in case mini-batches are used, due to the need to avoid having batches containing highly correlated samples that would slow down (or, in many cases, compromise) the performance of the model. In our model, we shuffle the training data before each training epoch, as well as the validation data before each validation. \\
\textbf{Plot.} The ``plot'' option in Matlab provides several pieces of information to be taken into account during the training process. Information include, but are not limited to, the mini-batch training loss and accuracy, the smoothed training loss and accuracy (i.e., the result of the application of a smoothing algorithm to the training accuracy), the validation loss and accuracy, hardware resources, etc. \\
\textbf{Validation Data.} The validation data, also known as validation set, refers to a subset of samples separated from the training set, that the model will rely on to evaluate the effectiveness of its training. In our case, by following the 80/20 rule, the validation set is represented by 20\% of the whole dataset. \\
\textbf{Validation Frequency.} The validation frequency represents the number of iterations between evaluations of validation metrics. We empirically set this value to $\lfloor \frac{|training\_set|}{miniBatchSize}\rfloor$.

\section{Performance}
\label{sec:performance}

\subsection{Category of Gun Identification}
\begin{figure}[h]
    \centering
    \includegraphics[width=\columnwidth]{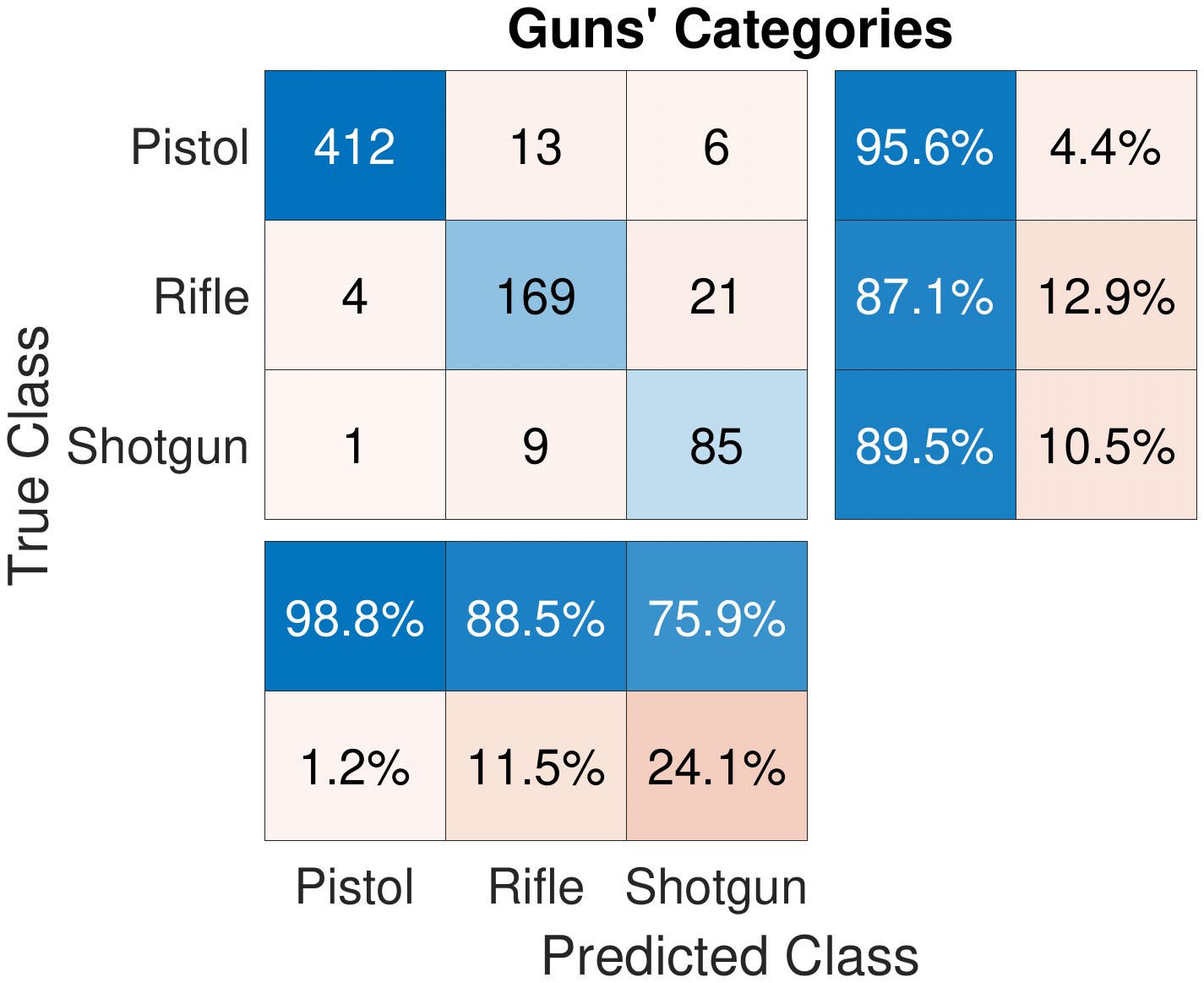}
    \caption{Confusion matrix associated with the classification of the Category of Guns.}
    \label{fig:category_performance}
\end{figure}

In this section, we consider the neural network previously introduced to infer the \emph{Category of gun}. We reconsider Table~\ref{table:dataset} and we divide the dataset into three classes, namely, \emph{Pistols}, \emph{Rifles}, and \emph{Shotguns}, according to the gun models in the dataset. Figure~\ref{fig:category_performance} shows the confusion matrix computed as the average of 50 training and validation runs.

The accuracy $acc$ can be computed according to Eq.~\ref{eq:accuracy}.

\begin{equation}
    acc = \frac{1}{N} \sum_{i=1}^{N_C} x_{ii}   
    \label{eq:accuracy}
\end{equation}

where $N=720$ is the total number of samples, $N_C = 3$ is the number of classes, and $x_{ii}$ is the $i^{th}$ diagonal element of the confusion matrix, yielding to $acc \approx 0.92$. The confusion matrix in Fig.~\ref{fig:category_performance} also reports summaries of columns and rows, predicted and true classes, respectively. 

We observe that the classification error spans between 4.4\% and 12.9\% for Pistol and Rifle classes, respectively. The class Rifle (an actual gunshot from a rifle) is incorrectly classified as either Pistol (4 times) or Shotgun (21 times) in the 12.9\% of the cases. The same type of analysis can be performed column-wise, where the prediction error spans between 1.2\% and 24.1\%. As an example, we observe that a prediction on class Shotgun is wrong in the 24.1\% of the cases (6 times for Pistol and 21 times for Rifle).

Finally, we observe that while the Pistol class is likely to be correctly classified all the times, the vast majority of errors are happening between the Rifle and Shotgun classes. 

\subsection{Caliber Identification}

\begin{figure}[h]
    \centering
    \includegraphics[width=\columnwidth]{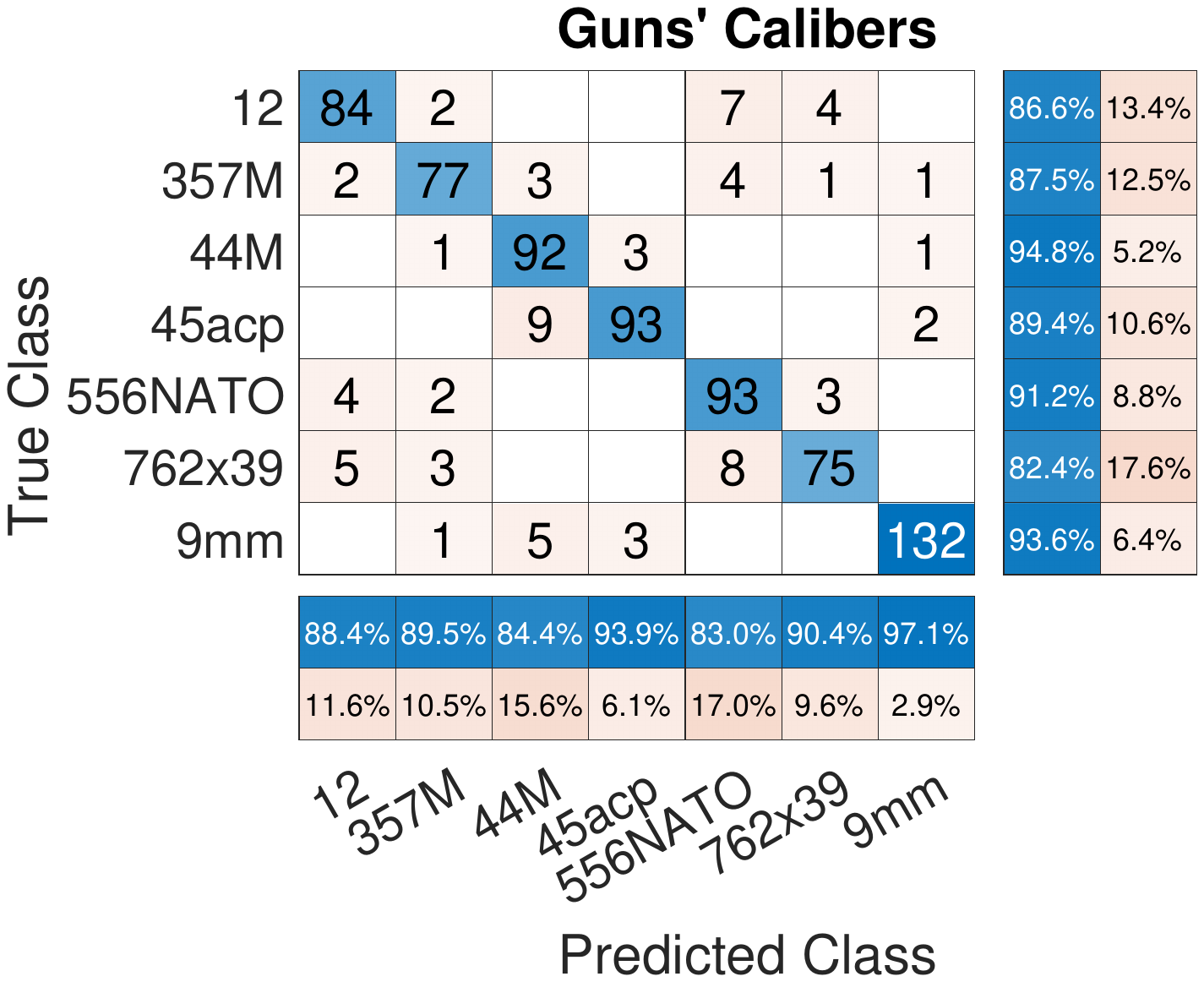}
    \caption{Confusion matrix associated with the classification of the Calibers of Guns.}
    \label{fig:caliber_performance}
\end{figure}

In this section, we report the performance of our classification algorithm when considering 7 different calibers from Table~\ref{table:dataset}. We group the video chunks based on gun caliber, obtaining 7 different classes, namely, 12, 357M, 44M, 45acp, 556NATO, 762x39, and 9mm. Figure~\ref{fig:caliber_performance} shows the confusion matrix computed as the average of 50 training and validation runs. The overall accuracy computed according to Eq.~\ref{eq:accuracy} sums up to $acc \approx 0.9$. Best and worst performance are achieved by 9mm and 762x39, respectively. In particular, class 762x39 is wrongly predicted 8 times as class 556NATO. Classes 556NATO and 762x39 are intrinsically similar, since both are from class Rifle. Therefore, they are prone to be confused. Nevertheless, we observe that this phenomenon is very limited since we have 3 cases of 556NATO classified as 762x39, and 8 cases for the opposite configuration. We also observe that 556NATO and 762x39 classes experience a significant amount of misclassifications with classes 12 and 357M. Conversely, class 44M and 9mm are the most likely to be correctly classified with 94.8\% and 93.6\%, respectively.

{\bf Highlights.} By combining Fig.~\ref{fig:category_performance} and Fig.~\ref{fig:caliber_performance} we can draw some interesting observations. Rifle class is misclassified for class Shotgun 21 times (the opposite is happening 9 times) in Fig.~\ref{fig:category_performance}, while 7+4=11 times (4+5=9 times) in Fig.~\ref{fig:caliber_performance}. We think that the error is not due to a specific caliber, either the 556NATO or the 762x39, but to the feature similarities between the two classes: Shotgun and Rifle classes. 

The pistol class is also misclassified as Rifle class 15 times. By looking into the details of Fig.~\ref{fig:caliber_performance}, we observe that the major source of misclassifications is coming from the 357M class, classified 4 times as 556NATO and 1 time as 762x39. We observe that the 357M is the most powerful among the pistol calibers hence it is the closest to Rifle class in terms of bullet size, pressure, and barrel diameter.

Finally, we observe that our solution is particularly robust in detecting pistols. In particular, one of the most adopted worldwide caliber (9mm) is characterized by a very limited number of misclassifications (9 out of 167 total). The same considerations apply to classes 44M and 45acp.

\subsection{Model Identification}

In this section, we consider all of the gun models previously introduced in Table~\ref{table:dataset} with the aim of classifying each of them. The total number of classes sums up to 59, which is the number of gun models considered throughout this paper. We report the confusion matrix associated with the aforementioned classification in Appendix~\ref{fig:gun_performance}. The accuracy sums up to $acc \approx 0.90$ and the maximum number of misclassifications (per model) never exceeds 2. We observe that class 38 (Ruger GP100 Match Champion) is never correctly classified. Finally, we highlight that the number of samples for the validation process is small (20\% of each gun model in Table~\ref{table:dataset}). Nevertheless, the diagonal of the matrix in Appendix~\ref{fig:gun_performance} collects the vast majority of the samples confirming the effectiveness of our model. We are confident that a larger data sample can increase the accuracy performance and effectiveness of gun model detection from gunshot sounds.  

\subsection{Testing}

To validate our methodology, we tested the model against a new set of audio samples taken from videos different than the ones considered before with varying conditions, including the background noise and relative positions between the microphone and the shooter. We consider a total of 115 audio samples constituted by 13 Pistols (Beretta 92 FS), 59 Rifles (Ruger AR, Daniel Defense M4 A1 SOCOM, Maadi AK-47) and 44 Shotguns (Maverick 88, Winchester Model 300 Defender). We observe that Pistol and Rifle classification is characterized by high performance, where only 4 Rifles samples are misclassified for Pistol. As for the Shotgun class, we highlight that the two shotguns considered are not in the training set (Table~\ref{table:dataset}) because we did not find any valid samples from additional videos. 
Although the audio samples are coming from different shotgun models, our algorithm can still detect the caliber with high probability (only 8 audio samples are misclassified), which verifies the effectiveness and correctness of our algorithm. Finally, we observe that the overall accuracy is consistent with the validation process and sums up to about 0.9.

\begin{figure}[h]
    \centering
    \includegraphics[width=\columnwidth]{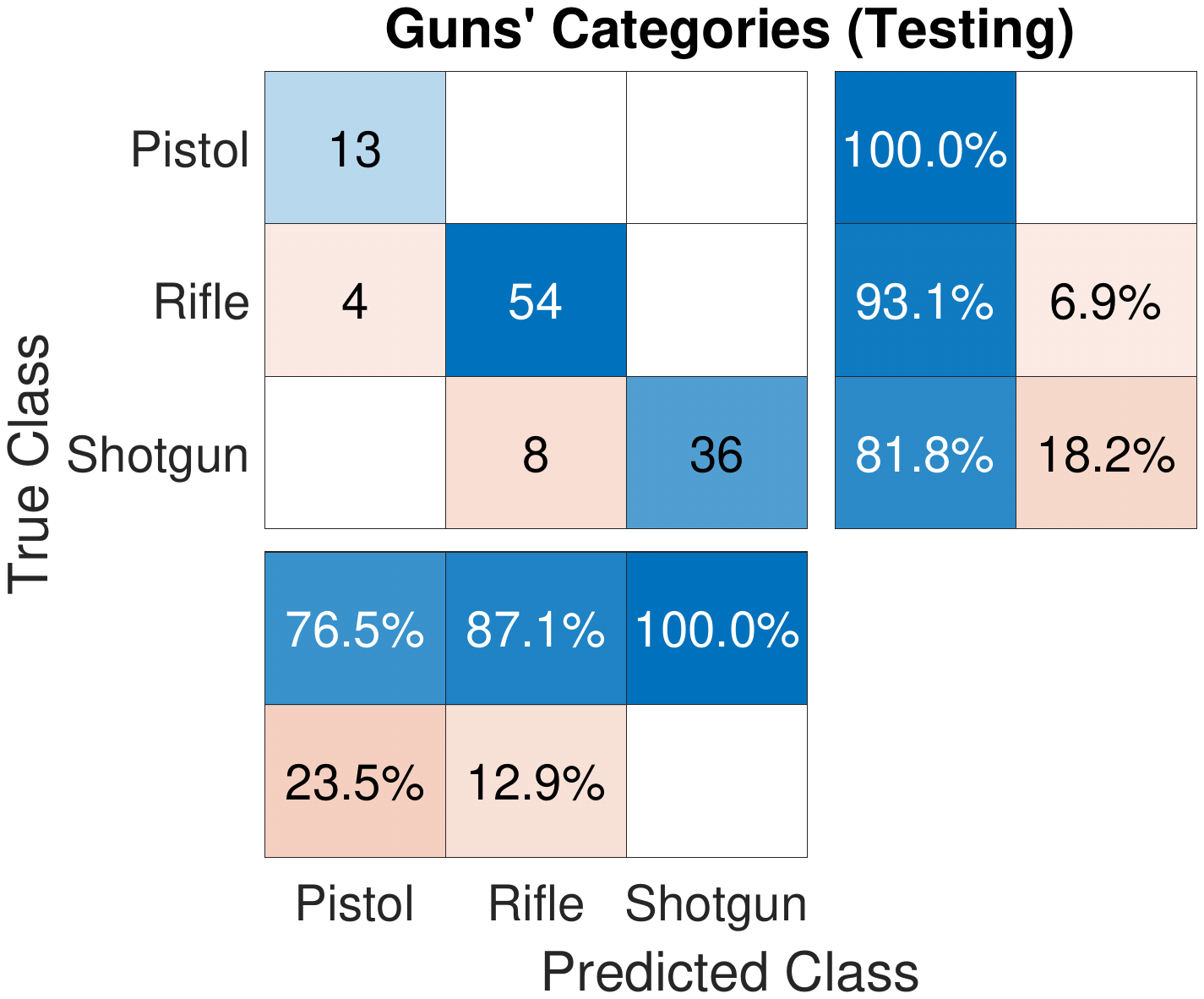}
    \caption{Confusion matrix associated with guns' categories testing while adopting new audio samples.}
    \label{fig:testing}
\end{figure}

\section{Conclusion}
\label{sec:conclusion}

Although scenarios requiring in-depth digital forensic of gunshots are countless, including military operations, mass-shooting, and possibly others, current solutions are far from reaching an adequate accuracy under real conditions. 

In this paper, we have proposed an effective and efficient methodology to uniquely fingerprint gunshots enabling the identification of the category, caliber, and the model of the gun with an accuracy higher than 90\% regardless of the capture conditions. 
Unlike existing solutions, our technique requires neither ad-hoc deployment of microphone networks, nor specific sample quality, and is agnostic to the microphone position with respect to the shooter. 
We have demonstrated that forensic analysis in the time-frequency domain of a single gunshot audio sample recorded by a commercial microphone (44100 samples per seconds) can be effectively used to infer the gun model (and other related characteristics). 
The proposed solution may lead to new insights and further developments in the area of weapon classification considering more samples, different noise levels, and a much larger weapon database.

\section*{Acknowledgment}
This publication was partially supported by the following awards: NPRP12S-0125-190013 from QNRF-Qatar National Research Fund, a member of The Qatar Foundation, and the Innovation Research Project ``Anti-Spoofers - GPS Spoofing Detection Exploiting Crowd-Sourced Information'' awarded by the HBKU Innovation Center. The information and views set out in this publication are those of the authors and do not necessarily reflect the official opinion of the QNRF.

\bibliographystyle{IEEEtran}
\balance
\bibliography{soundOfGuns}

\captionsetup{type=table}
\captionsetup{name=\textbf{Appendix}}

\begin{table*}[htbp]
\renewcommand{\thetable}{\textbf{\Alph{table}}}
\addtocounter{table}{-4}
\caption{Making up layers of our architecture.}
\vspace*{30mm}
\label{table:makingup-layers}
\centering
\begin{tabular}{|c|c|c|c|}
\hline
\textbf{Name} & \textbf{Type} & \textbf{Learnables} & \textbf{Total Learnables} \\ \hline
\textbf{inputImage} & \multirow{2}{*}{Image input} & \multirow{2}{*}{-} & \multirow{2}{*}{0} \\
(66x100x1 with 'zerocenter' normalization) & & & \\ \hline
\textbf{conv1} & \multirow{2}{*}{Convolution} & Weights: 3x3x1x40 & \multirow{2}{*}{400} \\
(40 3x3x1 convs, stride: [1 1], padding: 'same') & & Bias: 1x1x40 & \\ \hline
\textbf{batchnorm1} & \multirow{2}{*}{Batch Normalization} & Offset: 1x1x40 & \multirow{2}{*}{80} \\ (Batch Normalization, 40 channels) & & Scale: 1x1x40 & \\ \hline
\textbf{relu1} & \ac{ReLU} & - & 0 \\ \hline
\textbf{maxpool1} & \multirow{2}{*}{Max Pooling} & \multirow{2}{*}{-} & \multirow{2}{*}{0} \\
(3x3 Max Pooling, stride: [2 2], padding: 'same') & & & \\ \hline
\textbf{conv2} & \multirow{2}{*}{Convolution} & Weights: 3x3x40x80 & \multirow{2}{*}{28880} \\
(80 3x3x40 convs, stride: [1 1], padding: 'same') & & Bias: 1x1x80 & \\ \hline
\textbf{batchnorm2} & \multirow{2}{*}{Batch Normalization} & Offset: 1x1x80 & \multirow{2}{*}{160} \\
(Batch Normalization, 80 channels) & & Scale: 1x1x80 & \\ \hline
\textbf{relu2} & \ac{ReLU} & - & 0 \\ \hline
\textbf{maxpool2} & \multirow{2}{*}{Max Pooling} & \multirow{2}{*}{-} & \multirow{2}{*}{0} \\
(3x3 Max Pooling, stride: [2 2], padding: 'same') & & & \\ \hline
\textbf{conv3} & \multirow{2}{*}{Convolution} & Weights: 3x3x80x160 & \multirow{2}{*}{115360} \\
(160 3x3x80 convs, stride: [1 1], padding: 'same') & & Bias: 1x1x160 & \\ \hline
\textbf{batchnorm3} & \multirow{2}{*}{Batch Normalization} & Offset: 1x1x160 & \multirow{2}{*}{320} \\
(Batch Normalization: 160 channels) & & Scale: 1x1x160 & \\ \hline
\textbf{relu3} & \ac{ReLU} & - & 0 \\ \hline
\textbf{maxpool3} & \multirow{2}{*}{Max Pooling} & \multirow{2}{*}{-} & \multirow{2}{*}{0} \\
(3x3 Max Pooling, stride: [2 2], padding: 'same' & & & \\ \hline 
\textbf{conv4} & \multirow{2}{*}{Convolution} & Weights: 3x3x160x160 & \multirow{2}{*}{230560} \\
(160 3x3x160 convs, stride [1 1], padding: 'same') & & Bias: 1x1x160 & \\ \hline
\textbf{batchnorm4} & \multirow{2}{*}{Batch Normalization} & Offset: 1x1x160 & \multirow{2}{*}{320}      \\
(Batch Normalization, 160 channels) & & Scale: 1x1x160 & \\ \hline
\textbf{relu4} & \ac{ReLU} & - & 0 \\ \hline
\textbf{maxpool4} & \multirow{2}{*}{Max Pooling} & \multirow{2}{*}{-} & \multirow{2}{*}{0} \\
(1x13 Max Pooling, stride: [1 1], padding: [0 0 0 0] & & & \\ \hline
\textbf{Dropout (20\%)} & Dropout & - & 0 \\ \hline
\textbf{fc5} & \multirow{2}{*}{Fully Connected} & Weights: 7x1440 & \multirow{2}{*}{10087} \\
(7-neuron fully connected layer) & & Bias: 7x1 & \\ \hline
\textbf{Softmax} & Softmax & - & 0 \\ \hline
\end{tabular}
\end{table*}

\captionsetup{type=figure}
\captionsetup{name=\textbf{Appendix}}

\begin{figure*}[h]
    \renewcommand{\thefigure}{\textbf{\Alph{figure}}}
    \addtocounter{figure}{-12}
    \centering
    \caption{Confusion matrix associated with gun model classification.}
    \includegraphics[width=1.4\textwidth,angle=90]{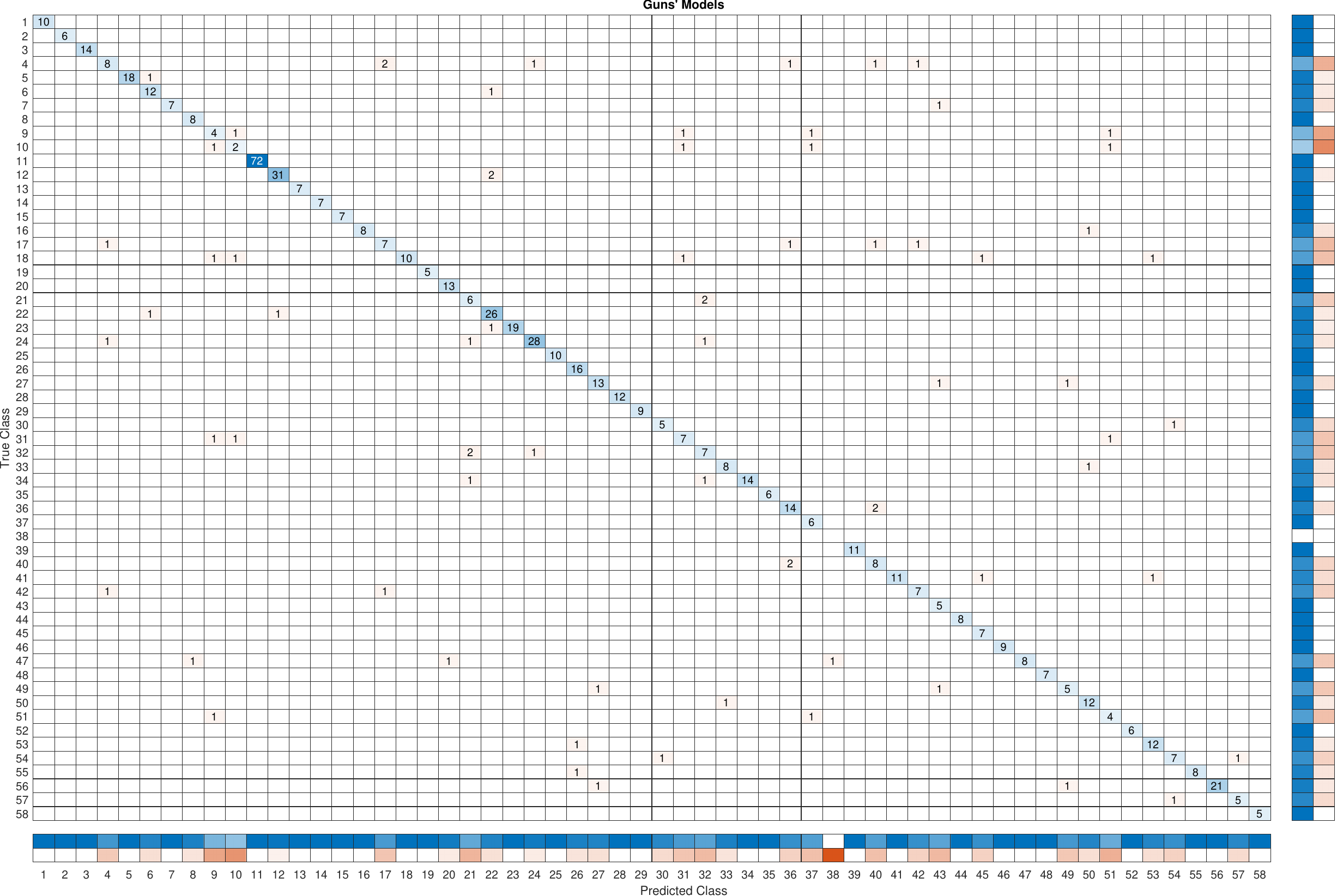}
    \label{fig:gun_performance}
\end{figure*}

\end{document}